\documentclass[11pt]{article}    
\usepackage[letterpaper,margin=1in]{geometry}

\usepackage{cite} % Make references as [1-4], not [1,2,3,4]
\usepackage{url}  % Formatting web addresses  
\usepackage{ifthen}  % Conditional 
\usepackage{multicol}   %Columns
\usepackage[utf8]{inputenc} %unicode support
\usepackage{authblk}
\usepackage{slashbox}
\usepackage{graphicx}
\usepackage{amsmath,amsfonts,amssymb}
\usepackage{epsfig}
\usepackage{enumerate}
\usepackage{multirow}
\usepackage{algorithm,algorithmic}
\usepackage{bm}
\usepackage{hhline}% http://ctan.org/pkg/hhline
\usepackage{array}
\newcolumntype{k}[1]{%
>{\raggedleft\hspace{0pt}}p{#1}}%
\newcolumntype{x}[1]{%
>{\centering\hspace{0pt}}p{#1}}%

\newcommand{\eq}[1]{(\ref{#1})}
\graphicspath{{figs//}}

\newcommand{\hide}[1]{}
\newcommand{\highlight}[1]{ #1 }

\usepackage{xcolor,colortbl}
\newcommand{\greencell}{\cellcolor{green}}

\begin{document}

\title{Reconstructing subclonal composition and evolution from whole
  genome sequencing of tumors}
 
\author[1]{Amit G. Deshwar}
\author[2]{Shankar Vembu}
\author[3]{Christina K. Yung}
\author[3]{Gun Ho Jang}
\author[3,4]{Lincoln Stein}
\author[1,2,4,5,6]{Quaid Morris}
\affil[1]{Edward S. Rogers Sr. Department of Electrical and Computer Engineering, University of Toronto}
\affil[2]{Donnelly Center for Cellular and Biomolecular Research, University of Toronto}
\affil[3]{Ontario Institute for Cancer Research}
\affil[4]{Department of Molecular Genetics, University of Toronto}
\affil[5]{Banting and Best Department of Medical Research, University of Toronto}
\affil[6]{Department of Computer Science, University of Toronto}
\date{}

\maketitle

\begin{abstract}      
Tumors often contain multiple subpopulations
of cancerous cells defined by distinct somatic mutations. We describe a new method, PhyloWGS, that can be applied to
WGS data from one or more tumor samples to reconstruct complete
genotypes of these subpopulations based on variant allele frequencies (VAFs)
of point mutations and population frequencies of structural variations.
We introduce a principled phylogenic correction for VAFs in loci affected by copy
number alterations and we show that this correction greatly improves subclonal
reconstruction compared to existing methods. 

\end{abstract}

%%%%%%%%%%%%%%%%%%%%%%%%%%%%%%%%%%%%%%%%%%%%%%%%%%%%%%%%%%%%%%
%%%%%%%%%%%%%%%%%%%%%% BACKGROUND %%%%%%%%%%%%%%%%%%%%%%%%%%%%%%
\section{Introduction}
Tumors contain multiple, 
genetically diverse subclonal populations of cells
that have evolved from a single progenitor population through
successive waves of expansion and selection \cite{nowell1976,Gerlinger12,hughes2014clonal}.
Reconstructing their evolutionary histories can help identify characteristic
driver mutations associated with cancer development
and progression \cite{Hanahan00,Hanahan11}; and can provide insight
into how tumors might respond to treatment \cite{aparicio2013implications,bedard2013tumour}.
In some cases, it is possible to genotype the subpopulations present
in a tumor, while reconstructing its history, using the population
frequencies of mutations that distinguish these subclonal
populations
\cite{Mullighan08,Navin10,Marusyk10,Gerlinger12,Schuh12,Shah12,Carter13,Landau13,theta,trap,phylosub,pyclone,expands,somatica,purbayes}.
Increasingly, tumors are being characterized using
whole genome sequencing (WGS) of bulk tumor samples \cite{pcawg} and few automated methods exist to reliably
perform this reconstruction on the basis of these data. 

Subclonal reconstruction algorithms attempt to infer the population
structure of heterogeneous tumors based on the measured variant
allelic frequency (VAF) of their somatic mutations.
Some methods perform this reconstruction based solely on single
nucleotide variants or small indels (collectively known as \emph{simple somatic mutations} or SSMs)
\cite{trap,pyclone,phylosub,expands,ding2012clonal,purbayes}.
Others use changes in read coverage to identify genomic regions with an average
ploidy that differs from normal which they explain using inferred copy
number variations (CNVs) affecting some of the cells in the sample
\cite{theta,absolute,cnanorm,somatica}.

The low read depth of current WGS complicates subclonal
reconstruction.
Until recently, subclonal populations (i.e., \emph{subpopulations})
were defined based on accurate estimates of
the proportion of cells with each mutation (i.e., their \emph{population frequency}) which, for individual SSMs, are
only available through targeted
resequencing where the read depths are orders of magnitude higher than
typical WGS depths\cite{phylosub,pyclone,ding2012clonal}.
However, preliminary
evidence suggests that the much larger number of mutations detected by
WGS can compensate for their decreased read depth \cite{nik12}. 
In contrast, CNVs affect large, multi-kilobase or megabase-sized
regions of the genome allowing the average
copy number of these regions to be accurately estimated with WGS.
Unfortunately, CNV-based subclonal reconstruction is more difficult
than SSM-based reconstruction because of the need to simultaneously
estimate population frequency and new copy number for each CNV.
Most CNV-based methods only attempting to infer the copy number status of the \emph{clonal} cancerous population
\cite{absolute,cnanorm} that contains the mutations shared by all of
the cancerous cells.
The few CNV-based methods \cite{theta, somatica} that attempt to resolve more than one cancerous
subpopulation are practically limited to a small number (often two) of subpopulations.
In contrast, SSM-based methods applied to targeted resequencing data
can reliably resolve many more cancerous
subpopulations \cite{phylosub, pyclone, trap, ding2012clonal}.
However, it remains unclear what the limits of WGS-based automated
subclonal reconstruction are.

Another open question is how to combine CNVs and SSMs when doing reconstruction.
CNVs overlapping SSMs can interfere with SSM-based
reconstruction because they complicate the relationship between VAF
and population frequency. 
Although some methods attempt to model the impact of CNVs on the
allele frequency of overlapping SSMs
\cite{expands, pyclone, phylosub, cloneHD}, these methods have
significant restrictions. For example, several of these methods
\cite{pyclone,phylosub} make the unrealistic assumption that every
cell either contains the structural variation and the mutation or
neither. Also, no method places structural variations in a phylogenetic tree, which is important for studying the evolution of cancerous genomes.  

We describe PhyloWGS, the first method designed for complete subclonal
phylogenic reconstruction of both CNVs and SSMs from whole genome
sequencing of bulk tumor samples.
Unlike all previous methods, PhyloWGS appropriately corrects SSM population
frequencies in regions overlapping CNVs and is fast enough to perform
reconstruction of at least five cancerous subpopulations based on thousands of mutations.
We present results on subclonal reconstruction problems that cannot be
correctly reconstructed using previous methods.
\highlight{We also probe the relationship between WGS read depth and the
  number of subpopulations that PhyloWGS can recover.} 
Finally, we demonstrate that even in the absence of reliable CNV estimates, it
is still feasible to perform automated subclonal composition reconstruction based
on SSM frequency data at typical WGS read depths (30-50x), even for
highly rearranged genomes where less than $2\%$ of the SSMs lie in
regions of normal copy number.

%Open source, free software implementing PhyloWGS is available here:
%\textrm{http://morrislab.med.utoronto.ca/phylowgs/}.

\section{Previous work}

\begin{figure}[!t]
\centering
\includegraphics[scale=.225]{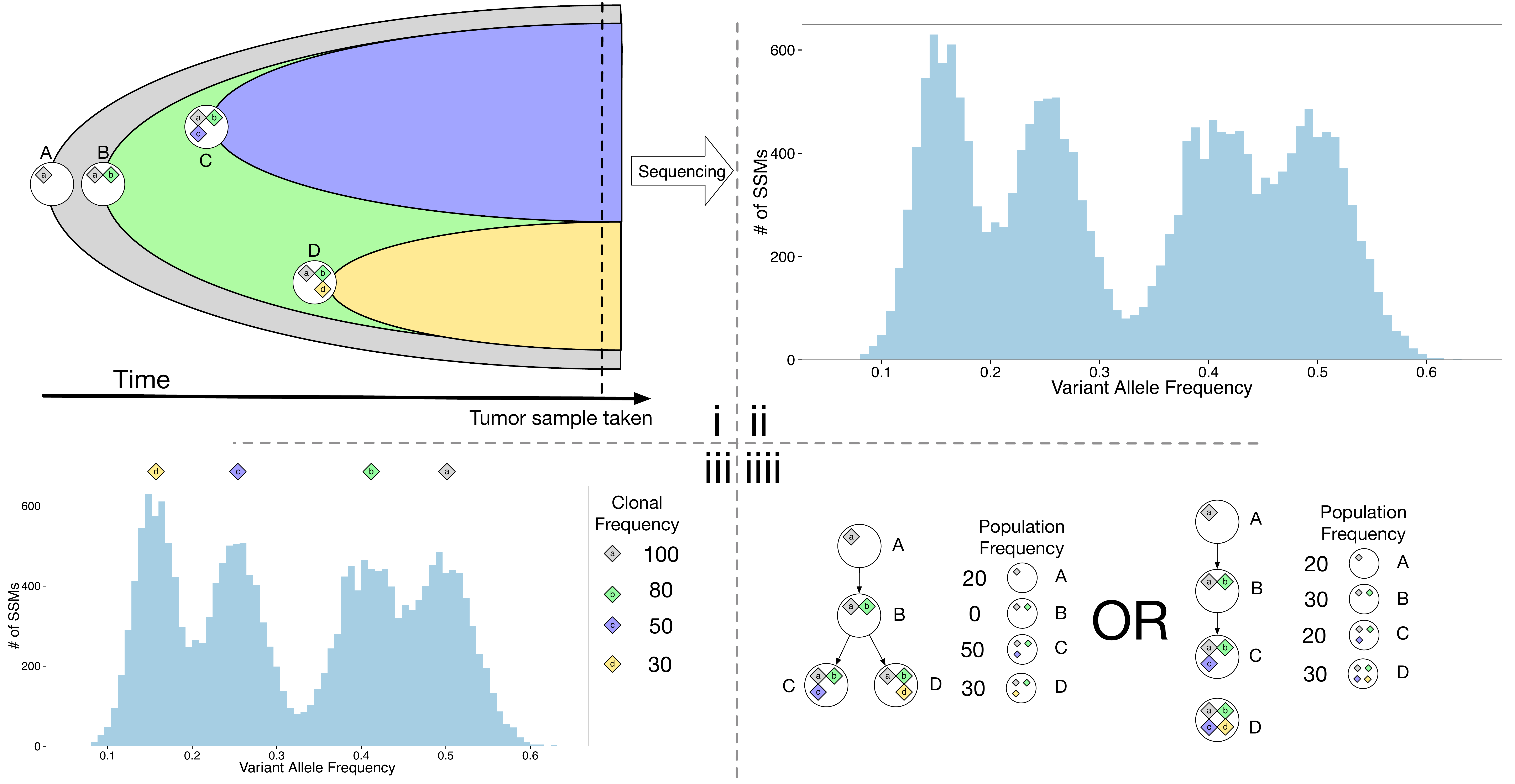}
  \caption{{\bf The development of intratumor heterogeneity and subclonal reconstruction.} Tumor composition over time (i), the resulting distribution of variant allele frequencies (ii), the result of successful inference of the VAF clusters (iii), and the desired output of subclonal inference (iiii)}
  \label{fig:clonal}
\end{figure}

Figure \ref{fig:clonal} provides an overview of an
evolving tumor, measuring of VAFs, and the resulting subclonal
reconstruction process.
Panel (i) of this figure shows a visualization of the evolution of a
tumor over time as noncancerous cells (subpopulation A, shown in grey) are replaced by, at
first, one clonal cancerous population (subpopulation B, shown in green) which then further
develops into multiple cancerous subpopulations (C and D, shown in
blue and yellow, respectively).
Tumor cells define new subpopulations by acquiring new oncogenic
mutations that allow their descendants to expand relative to the
other tumor subpopulations. Each circle in Panel (i) refers to a
subpopulation.
We associate subpopulations with the set of shared somatic
mutations that distinguish it from its parent subpopulation (or, in
the case of A, from the germline (or reference) genome); this
mutation set is indicated by the corresponding lower case letter
(e.g. mutation set \emph{b} first appears in subpopulation \emph{B}).
However, each subpopulation also inherits all of its parent's
mutations; the \emph{subclonal lineage} of a mutation
is the set of all subpopulations that contain it (e.g., the subclonal
lineage of a is A, B, C, and D).

In general, the subpopulation-defining mutation sets include more than one mutation.
Cancerous cells often have increased
mutation rates, and even noncancerous cells accumulate
somatic mutations at a rate of $1.1$ per cell division \cite{behjati2014genome}. 
As such, subpopulations are defined not only by the small number of oncogenic
`driver' mutations that support rapid expansion but also a larger number
of `passenger' mutations acquired before the driver mutation(s).
The selective sweeps that cause subpopulation expansion increase the population frequency of
both driver and passenger SSMs, driving them to having
indistinguishable population frequencies \cite{burrell2013causes,klein2013selection}.
However, sampling and technical noise in sequencing means that the observed VAFs are distributed around the true value for a subpopulation. Panel (ii) shows an example histogram of measured
VAFs for SSMs found in a heterogeneous tumor sample.

Subclonal reconstruction algorithms define mutation sets, and their associated subpopulations, by
analyzing the population frequencies of somatic mutations detected in a tumor sample.
In Figure \ref{fig:clonal}, all mutations are SSMs,
and all SSMs occur on one copy in diploid regions of the
genome. In this case, the estimated population frequency of an SSM is simply twice its VAF.
Figure \ref{fig:cnv_ex}, discussed in the next section, shows
how CNVs overlapping SSM loci change this relationship.
Note that although each VAF cluster corresponds to a subclonal
lineage, and a subpopulation that was present at some point during the
tumor's evolution, this subpopulation need not be present when the
tumor was sampled. In Figure \ref{fig:clonal},
subpopulation B is no longer present in the tumor, although its two
descendant subpopulations are.
These \emph{vestigial} VAF clusters, if they exist, always correspond to subpopulations at
branchpoints in the phylogeny, however, not every branchpoint
generates a vestigial cluster.

\subsection{SSM-based approaches}
SSM-based subclonal reconstruction algorithms attempt to reconstruct the
subpopulation genotypes based on VAF clusters (and their associated
mutation sets) identified by
fitting statistical mixture models to the
VAF data either without phylogenic reconstruction \cite{pyclone,sciclone,expands,purbayes}, before phylogenic reconstruction\cite{recBTP} or 
concurrently with it\cite{trap,phylosub}.   
Often, as in Figure \ref{fig:clonal}, the clusters are overlapping which introduces
uncertainty in the exact number of mutation sets 
represented in the tumor (as well as in the assignment
of SSMs to clusters).
Adding more clusters to the model always provides better data fit, so to
prevent overfitting, the cluster number is selected by
balancing data fit versus a complexity penalty (e.g. the Bayesian Information Criteria) or by Bayesian inference in a
non-parametric model\cite{pyclone, phylosub,sciclone}.

In panel (iii) in Figure \ref{fig:clonal} the correct number of clusters have been
recovered along with appropriate central VAFs.

Assuming that the correct VAF clusters can be recovered, the subclonal
lineages corresponding to each mutation set must still be defined.
Defining the subclonal lineages is equivalent to defining the tumor
phylogeny; and often multiple phylogenies are consistent with the
recovered VAF clusters (e.g. panel (iiii) in Figure \ref{fig:clonal}).
Complete and correct reconstruction of subpopulation genotypes
requires resolving this ambiguity.
To do so, reconstruction methods make one of a handful of assumptions about the process of tumor evolution.

A common, and powerful, assumption is the infinite sites assumption (ISA) \cite{kimura69,hudson83,phylosub} 
which posits that each SSM only occurs once in the evolutionary history of the tumor.
The ISA implies that the tumor evolution is consistent with a `perfect and persistent
phylogeny' \cite{pyclone}: each subpopulation has all of the SSMs that its ancestors had;
each SSM appears in only one subclonal lineage; and each subclonal
lineage corresponds to a subtree in the phylogeny of
 tumor subpopulations.
Because SSMs are relatively rare (compared to the genome size), the
ISA is nearly always valid for all SSMs. So there is little danger of
incorrect reconstructions due to violations of the ISA.
In many cases, the ISA alone permits the recovery of multiple, complete subpopulation genotypes from a
single or small number of tumor samples using either the \emph{sum rule}
\cite{phylosub} (also called the pigeonhole principle \cite{nik12}) or the
\emph{crossing rule} \cite{phylosub}, respectively. 
Methods that do not use the ISA require, in the case of no measurement
noise, at least as many tumor
samples as there are subpopulations \cite{trap, clomial}; in actual
application when there is noise, even more samples are required.

Unfortunately, the ISA alone is often unable to fully resolve
reconstruction ambiguity.
As such, some methods \cite{recBTP, trap} also make a sparsity assumption to
select among ISA-respecting phylogenies consistent with the VAF frequency data.
This assumption, which we call \emph{strong parsimony}, posits that
due to expansion dynamics, there are a small number of subpopulations
still present in the tumor \cite{recBTP, trap}, and that many of the
VAF clusters are vestigial.
These methods therefore select the phylogeny (or phylogenies) that maximize the number of vestigial
VAF clusters \cite{trap}, or equivalently, the number of branchpoints
where the parental subpopulation has a zero frequency in the current
tumor\cite{recBTP, trap}.
The strong parsimony assumption does resolve some ambiguity, and leads
to the correct reconstruction in Figure \ref{fig:clonal}, but it is risky as its empirical validity is not yet established. 
For example, under some conditions, a linear (i.e. non-branching)
phylogeny can be mistaken for a branching one; the
risk of this occurring increases as the VAF measurement noise or the number
of subpopulations in the tumor increases.
This background distribution of false positive vestigiality is not yet
considered by either of the methods that assume strong parsimony.

By assigning all SSMs within a VAF cluster to the same
mutation set; reconstruction methods make another implicit assumption
which we call \emph{weak parsimony}.
This assumption does not hold if two mutation sets have the same
population frequency.
Note that if the ISA is valid, by the pigeonhole principle, weak
parsimony is guaranteed to be valid whenever the population frequency of the mutation set is $>50\%$.

Table \ref{tbl:methods} classifies reconstruction methods based
on these assumptions, whether they recover complete
subpopulation genotypes (or simply identify subclonal lineages), and
whether they can handle single tumor samples, multiple tumor
samples, or both.

\begin{table}[!t]
\caption{Table of subclonal reconstruction methods, their properties and assumptions}
\begin{center}
\begin{tabular}{|c|p{1.2cm}|p{1.1cm}|p{.8cm}|p{.9cm}|p{.7cm}|p{.8cm}|p{.9cm}|}
\backslashbox{Property}{Method} & PhyloWGS  & PhyloSub \cite{phylosub} & THetA \cite{theta} & PyClone \cite{pyclone} & TrAp \cite{trap} & Clomial \cite{clomial} & RecBTP \cite{recBTP}  \tabularnewline
\hline
SSM & \greencell Y	& \greencell Y & N & \greencell Y & \greencell Y & \greencell Y &  \greencell Y \tabularnewline
CNV &  \greencell Y	& N & \greencell Y & N & N & N & N \tabularnewline
Weak parsimony &	Y	& Y & N/A & Y & Y & Y & Y  \tabularnewline
Strong parsimony &	 \greencell N	& \greencell N & N/A &  \greencell N & Y &  \greencell N & Y \tabularnewline
Infinite sites &	Y	& Y & N/A & Y & Y & N & Y \tabularnewline
Phylogenetic inference & \greencell Y	&  \greencell Y & N & N & \greencell Y & N & \greencell Y\tabularnewline
Parametric &	\greencell N	& \greencell N & Y & \greencell N & \greencell N & Y & \greencell N \tabularnewline
Multiple samples &	\greencell Y	& \greencell Y & N &  \greencell Y & N & \greencell Y & N\tabularnewline
Genotype uncertainty&	\greencell Y	& \greencell Y & N & N/A & N & N & N \tabularnewline
\end{tabular}
\end{center}
\begin{flushleft}
\end{flushleft}
\label{tbl:methods}
\end{table}

PhyloWGS, like its predecessor PhyloSub\cite{phylosub}, does not make the strong
parsimony assumption nor does it only report a single tree.
Instead it reports samples from the posterior
distribution over phylogenies.
Because the clustering of the VAF frequencies is performed
concurrently with phylogenic reconstruction, PhyloWGS is
able to perform accurate reconstruction even when the weak parsimony assumption is violated in a strict subset of the samples available.
For example, if the VAF clusters overlap in one sample but
not another.
Our Markov Chain Monte Carlo (MCMC) procedure samples phylogenies
from the model posterior that are consistent with the mutation
frequencies and does not rule out phylogenies that are equally consistent with the data. From this collection of
samples, areas of certainty and uncertainty in the reconstruction can be
determined.

\subsection{CNV-based approaches}
There are three major differences between CNV-based reconstruction and
SSM-based reconstruction.
First, because large regions of the genome are affected by CNVs and
reads mapping across the regions can be used to estimate average ploidy, accurate
quantification of changes in average copy number can be achieved with
much smaller read depths (as low as 5-7x) \cite{bicseq, theta}.
However, the other two differences make CNV-based subclonal reconstruction more
difficult and less generally applicable compared with SSM-based methods.
One difference is that the ISA is often invalid because CNVs affect
large regions of the genome. As such, it is more common to see
overlapping mutations occurring in independent cells; these make the reconstruction
problem more challenging.
Even in the circumstance that only one CNV affects a given region, inferring its
population frequency is still challenging because at least two values,
population frequency $\phi \in (0, 1)$ and non-negative integer copy number $C$, have to be
simultaneously inferred from a single observed, non-normal average
copy number $x \neq 2$. In particular, this equation,
\[
  x = \phi C + (1-\phi) 2,
\]
always has at least two different solutions for $x>1$.

In absence of other information, like B-allele frequencies
\cite{nik12}, parsimony assumptions are relied upon to resolve
reconstruction ambiguities.
One strategy only attempts to reconstruct the cancerous, subclonal
lineage \cite{absolute, cnanorm} with the highest population frequency (also known as the `clonal population').
From this reconstruction, the proportion of cells in the tumor sample that are
cancerous (i.e. the cellularity), as well as detecting CNVs
that are shared by all cancerous cells in the tumor can be derived.
However, this approach can fail when there are multiple subclonal
populations, especially if they share few CNVs \cite{theta, somatica}.
Methods that attempt to detect $>1$ cancerous subpopulation do so by
balancing data fit with a complexity term that penalizes additional subpopulations\cite{theta,somatica}.
So far, these methods seem to be practically limited to a small number
of cancerous subpopulations (i.e., $2$); and cannot be applied to
tumors with substantial rearrangements.

\subsection{Combining SSMs and CNVs}
\begin{figure}[!t]
\centering
\includegraphics[scale=.5]{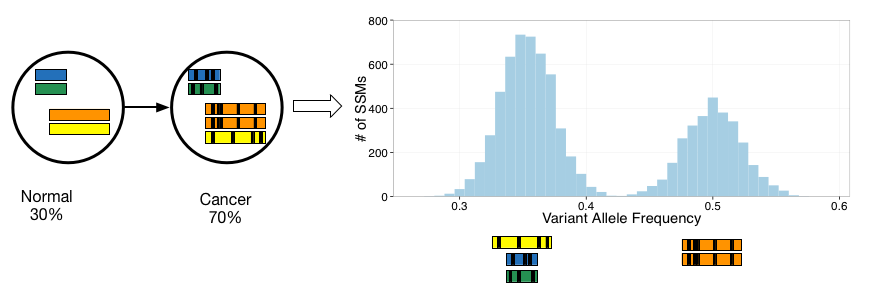}
  \caption{{\bf Example of CNVs affecting the distribution of VAFs.} }
  \label{fig:cnv_ex}
\end{figure}

\highlight{In loci affected by CNVs, computing the population frequency
of an SSM from its VAF requires knowing whether the SSM occurred
before, after or independently of the CNV. If the SSM occurred before the CNV, and CNV
affects the copy number of the SSM, then computing its VAF also requires knowing the new
number of maternal and paternal copies of the locus.}
Figure \ref{fig:cnv_ex} illustrates a situation where incorporating
CNV information is critical for subclonal reconstruction. Without CNV
information, the two VAF peaks would be interpreted as two separate
subclonal lineages. With CNVs, it becomes clear that the second peak is caused by the amplification of part of the genome that increases the VAF of all SSMs found in the region. 

Some subclonal reconstruction methods simply ignore the impact that
CNVs have on the relationship between SSM population and allele frequency \cite{trap, purbayes}.
Other methods that do account
for the effect of copy number changes on SSM frequencies
\cite{pyclone, phylosub, expands} by integrating over all the possible
relationships between allele frequency and population frequency
without making use of the fact that the infinite sites assumption,
which was necessary to uniquely associate SSMs to subclonal lineages in
the first place, also constrains this relationship \cite{nik12}. 

For the first time, we describe an automated method, PhyloWGS, that performs subclonal reconstruction using both CNVs and SSMs.
By combining information from both CNVs and SSMs, and properly accounting for their interaction, we provide a more comprehensive and accurate description of subclonal genotype.

%%%%%%%%%%%%%%%%%%%%%%%%%%%%%%%%%%%%%%%%%%%%%%%%%%%%%%%%%%%%%%%%
%%%%%%%%%%%%%%%%%%%%%%%%%%%%%%%%%%%%%%%%%%%%%%%%%%%%%%%%%%%%%%

%%%%%%%%%%%%%%%%%%%%%%%%%%%%%%%%%%%%%%%%%%%%%%%%%%%%%%%%%%%%%%
%%%%%%%%%%%%%%%%%%%%%% RESULTS %%%%%%%%%%%%%%%%%%%%%%%%%%%%%%
\section{Results}
In the following, we first provide a brief explanation of how PhyloWGS incorporates both SSMs and CNVs in phylogenic reconstruction by converting CNVs into pseudo-SSMs and performing subclonal reconstruction on the SSMs and pseudo-SSMs; full details are provided in the \textrm{Methods} section.
Then, we show an illustrative example where accounting for the effect of CNVs on SSMs permits the correct subclonal reconstruction of a tumor population whereas using either CNV or SSM data in isolation does not.
Then, we describe our efforts to quantify the relationship between
read depth and the number of subpopulations that can be accurately
recovered by applying PhyloWGS to simulated WGS data of different read
depths, number of subpopulations, and SSMs.
Next, we describe the application of PhyloWGS to three TCGA benchmark datasets.
\highlight{Finally, we describe the application of PhyloWGS to two real datasets: a multiple-sample WGS data from a patient with chronic lymphocytic leukemia and a single sample from a breast tumor.}

\subsection{Incorporating CNVs with SSMs in phylogenic reconstruction}
We assume that a CNV algorithm has already been applied to the sequencing data and that this algorithm 
provides estimates of copy number $C$ and population frequency $\phi_i$ for each CNV $i$. 
We use these estimates in two ways: first, for each CNV, we create an
equivalent {\it pseudo-SSM} with population frequency $\phi_i$ by
adding an SSM to the dataset with total reads $d_i$ and variant reads
$d_i * \phi_i / 2$ rounded to the nearest whole number (i.e., the
expected number of variant reads of a heterozygous mutation with
population frequency $\phi_i$) where $d_i$ is set to a user-defined
multiple of the average WGS read-depth.  If a confidence interval for
$\phi_i$ is available, we can set $d_i$ to have the same confidence
interval. \highlight{Note, we allow multiple CNVs to affect the same locus; each
of these CNVs is assigned its own pseudo-SSM.}

Second, we associate all SSMs within the region affected by the CNV to
this pseudo-SSM.  Our model (described in the \textrm{Methods}
section) uses this association to compute the transformation from the
inferred population frequency of an SSM to its expected VAF.

Here, we briefly describe how this transformation when there is only one CNV affecting the SSM locus; the
\textrm{Methods} section describes the general version of the
transformation, used by PhyloWGS, that allows multiple CNVs to
affect the locus.

Given the population frequency of the CNV, $\phi_c$, the copy number of the CNV $C$ broken down into maternal and paternal components $C = C^p + C^m$, and the population frequency of the SSM, $\phi_s$, the equations below compute the expected allele frequency of the SSM $x_{ssm}$. Here we are using the terms `maternal' and `paternal' simply to distinguish the two copies and not to suggest that we have actually assigned each chromosome to each parent. 
Furthermore, the description here assumes that the SSM is on the maternal copy, if it is on the paternal copy, replace $C^m$ with $C^p$ below.

If a SSM lies in a region affected by a CNV, there are three possibilities for their phylogenic relationship:

\begin{enumerate}[1)]
\item SSM precedes the CNV event, i.e., the CNV occurred in a cell already containing the SSM
\item SSM occurs after the CNV event, i.e., the SSM occurred in a cell already containing the CNV 
\item SSM and CNV occurred in separate branches of phylogeny, i.e., the mutations occur in separate cells and no cell contains both the SSM and the CNV.
\end{enumerate}

\subsubsection*{Case 1 (SSM $\rightarrow$ CNV)}
Because of the infinite sites assumptions, this phylogenic
relationship requires $\phi_s \geq \phi_c > 0$.
In this case, cells with the CNV contain $C^m$ copies of the SSM and cells with the SSM but not the CNV only have one mutated copy. As such, the expected allele frequency can be written as:
\begin{align*}
x_{ssm} = \frac{C^m \phi_c + (\phi_s-\phi_c)}{2(1-\phi_c) + C\phi_c}
\end{align*}
The numerator corresponds to the average number of copies per cell of the SSM-mutated locus in the population and the denominator is the average number of copies per cell of the locus (mutated or not) in the population.
We note that if there is no copy number change in $C^m$ then the
numerator is simply $\phi_s$; and if $C^m = 0$ then the numerator is $\phi_s - \phi_c$.

\subsubsection*{Case 2 (CNV $\rightarrow$ SSM)}
This case is only possible if the maternal locus still exists after
the CNV (i.e. $C^M \geq 1$), and furthermore that $\phi_c \geq \phi_s > 0$.
By the infinite sites assumption, only one copy of the locus is affected, so the numerator is simply $\phi_s$ and we do not need to know the breakdown of $C$ into $C^m$ and $C^p$. As such:
\begin{align*}
x_{ssm} = \frac{\phi_s}{2(1-\phi_c) + C\phi_c}
\end{align*}

\subsubsection*{Case 3 ($\genfrac{}{}{0pt}{}{\nearrow}{\searrow}^{SSM}_{CNV}$)}
In this case, the SSM and CNVs lie on different branches of the
phylogeny and no cell in the population contains both mutations, so
the only constraints on $\phi_s$ and $\phi_c$ are that $\phi_s +
\phi_c \leq 1$.
As per Case 2, the average number of loci affected by
the SSM is $\phi_s$. So the expected allele frequency is identical to
case 2.
\begin{align*}
x_{ssm} = \frac{\phi_s}{2(1-\phi_c) + C\phi_c}
\end{align*}

\begin{figure}[!t]
\centering
\includegraphics[scale=.5]{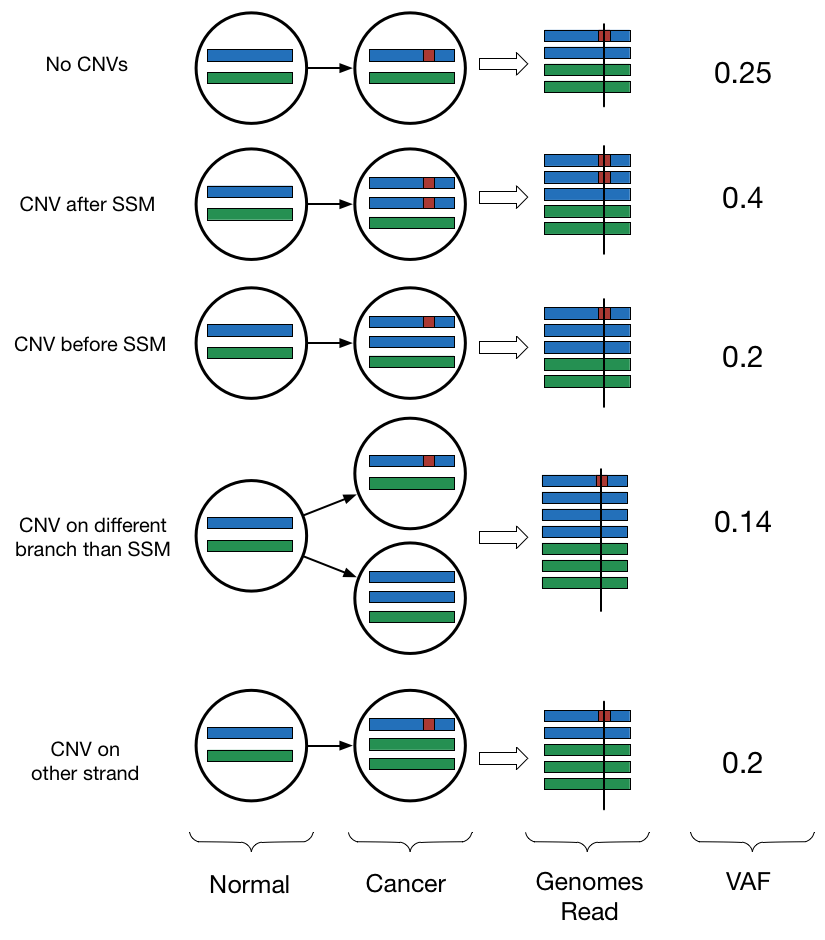}
  \caption{{\bf Changes to VAF caused by CNVs with different phylogenetic relationships.} }
  \label{fig:cnv_ssm}
\end{figure}

We illustrate some of the ways how the relationship between a CNV and an affected SSM in the phylogenetic tree affects the observed VAF of that SSM in Figure \ref{fig:cnv_ssm}. 

We note that the breakdown of $C$ into $C^m$ and $C^p$ and phasing the
SSM only affects the expected variant allele frequency in Case 1. This
is because it is the only case where a CNV event can affect a mutated
locus. Although PhyloWGS requires the breakdown of $C$ into $C^m$ and
$C^p$ under these conditions, we do not require the SSM to be phased,
as many cannot be\cite{nik12}, and instead consider both possibilities when computing the
likelihood. \highlight{Some subclonal copy number callers
  decompose $C$ into $C^m$ and $C^p$ \cite{titan}; if the caller does
  not provide this decomposition, then PhyloWGS should be run on loci
  where \hbox{$C \in \{0,1,2\}$.}}

An important consequence of these rules is that under some conditions,
it is possible to unambiguously identify a branching phylogeny using single-sample data.  If an SSM can be phased to an amplified locus there are situations where given particular values of $x_{ssm}$, $\phi_c$, $C^p$ and $C^m$ one can distinguish between Case 1 and Case 3.  For example, given $x_{ssm} = 0.1, \phi_c = 0.4, C^m=10, C^p=1$, for Case 3 the inferred $\phi_s$ is $0.56$.  However, if Case 1 were true the resulting inferred $\phi_s$ would be negative and so Case 1 is not possible.
This condition holds whenever $x_{ssm}*(2(1-\phi_c)+C \phi_c) < (C^m-1)\phi_c$.
We were unable to find any other circumstances in which single sample
VAFs were more consistent with a branching phylogeny than a chain
phylogeny.

\subsection{The combination of CNVs with SSMs is required for accurate subclonal reconstruction}

\begin{figure}[!t]
\centering
\includegraphics[scale=1]{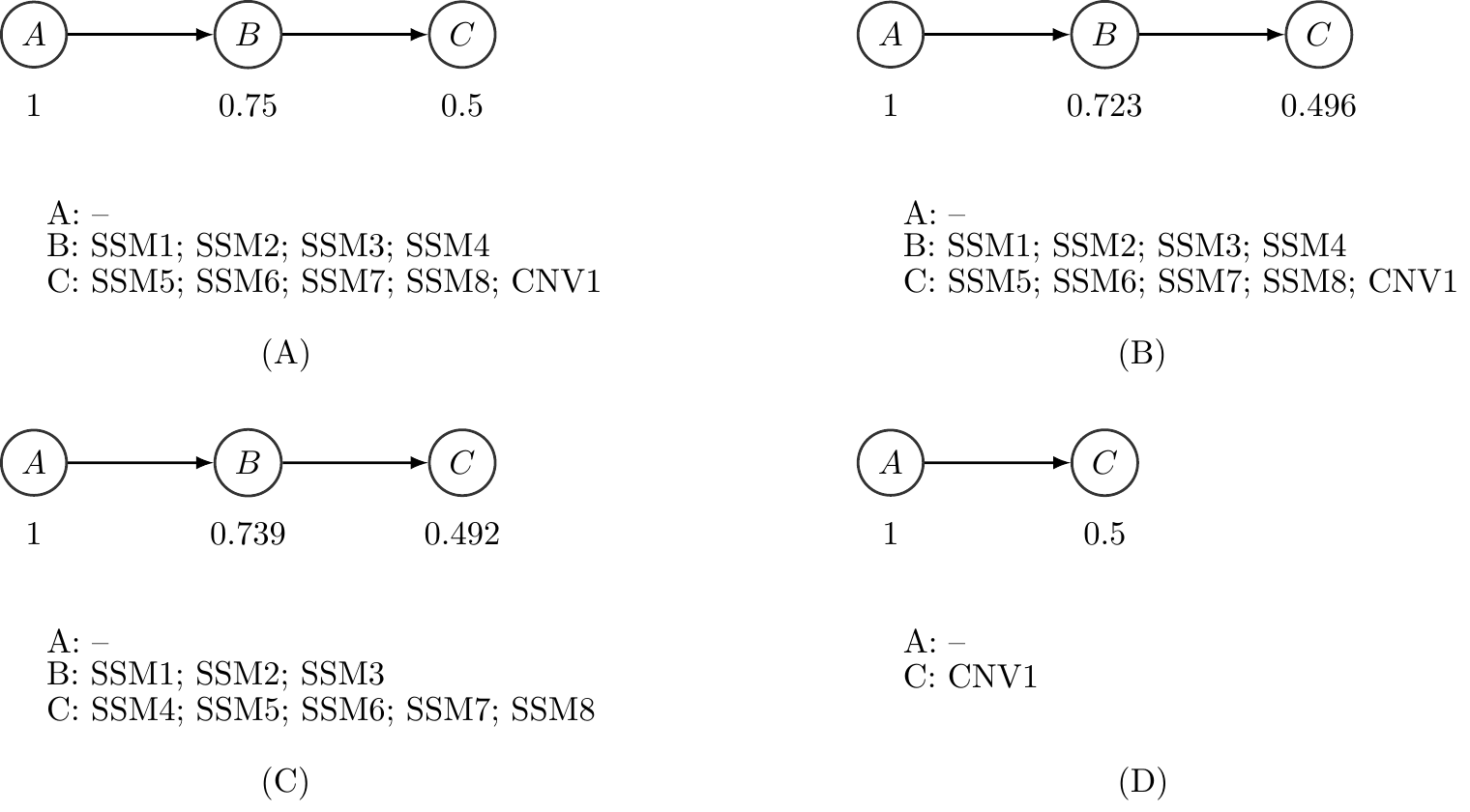}
  \caption{{\bf Example subclonal structure and inferred phylogenies using different methods.}
      A: Example of tumor subclonal structure.  B: Tumor phylogeny
      recovered by PhyloWGS.  C: Tumor phylogeny recovered by
      PhyloSub. D: Subclonal structure implied by only CNVs.}
\label{fig:tree}
\end{figure}

Consider a tumor where 25\% of the cells are normal (no SSMs and
diploid, population A), 25\% come from a subpopulation with only SSMs
(SSM1-4, population B) and 50\% belong to a descendant subpopulation
of B containing all the SSMs from B and adding new simple somatic
variants (SSM5-8) and a homozygous deletion (CNV1) in the region
containing SSM4, labelled population C. The evolutionary tree of this
population is shown in Figure \ref{fig:tree}A. 
In reads sampled from
this population, the expected variant allele frequencies for SSM1-3
are 37.5\% (i.e. half of their population frequency) and for SSM5-8
they are 25\%; however based on the rules described in the Methods
section, the expected variant allele frequency of SSM4 is 25\%.  This
is because all the copies of the genome at that position come from
population A or B.  Population A and B are present in equal
proportions and only one copy in population B contains variant reads,
so 25\% of the genomes contain the variant allele.  As such, methods
that do not incorporate the CNV change at the SSM4 locus will
incorrectly assign SSM4 to population C. Also, methods that
incorporate only CNV information cannot detect the subpopulation B
which is defined by SSM alone.

We generated simulated variant and reference allele counts for this example at a simulated read-depth of 60. A table containing the reference and total read counts for each SSM can be found as Table \ref{tbl:readcounts}. 
PhyloWGS was able to correctly reconstruct the evolutionary history and subpopulation structure (Figure \ref{fig:tree}B).  However, a version of PhyloSub which ignored CNVs incorrectly assigned SSM4 to population C (Figure \ref{fig:tree}C).  Furthermore, by construction, there is no way to recover population B based only on CNV data, so a perfect CNV-based algorithm would infer the subclonal structure in Figure \ref{fig:tree}D.
\begin{table}[!t]
\caption{Table showing reference and total read counts for example tumor sequencing data }
\begin{center}
\begin{tabular}{|l|c|c|}
mutation id & reference counts  & total counts \tabularnewline
\hline
s0 & 23	& 47 \tabularnewline
s1&	35&	69\tabularnewline
s2&	26&	51\tabularnewline
s3&	29&	56\tabularnewline
s4&	30&	40\tabularnewline
s5&	42&	70\tabularnewline
s6&	27&	41\tabularnewline
s7&	36&	56\tabularnewline
s8&	59&	75\tabularnewline
s9&	57&	76\tabularnewline
s10&	51&	68\tabularnewline
s11&	37&	51\tabularnewline
\end{tabular}
\end{center}
\begin{flushleft}
\end{flushleft}
\label{tbl:readcounts}
\end{table}

We also ran PyClone \cite{pyclone} on this dataset.  PyClone cannot take as input the fact that a locus has been homozygously deleted, so we ran PyClone either by telling it there were no CNV changes or that there was a deletion of one copy.  Without any input CNV alterations, PyClone produced a clustering identical to PhyloSub, while in the single copy deletion state, PyClone placed SSM4 in an additional cluster with no other mutations. 
As this simple example illustrates, integrating data from both SSMs and CNVs is required for full, and accurate, subclonal reconstruction.

\subsection{Applying PhyloWGS to simulated data}
\begin{figure}[!t]
\centering
\includegraphics[scale=.9]{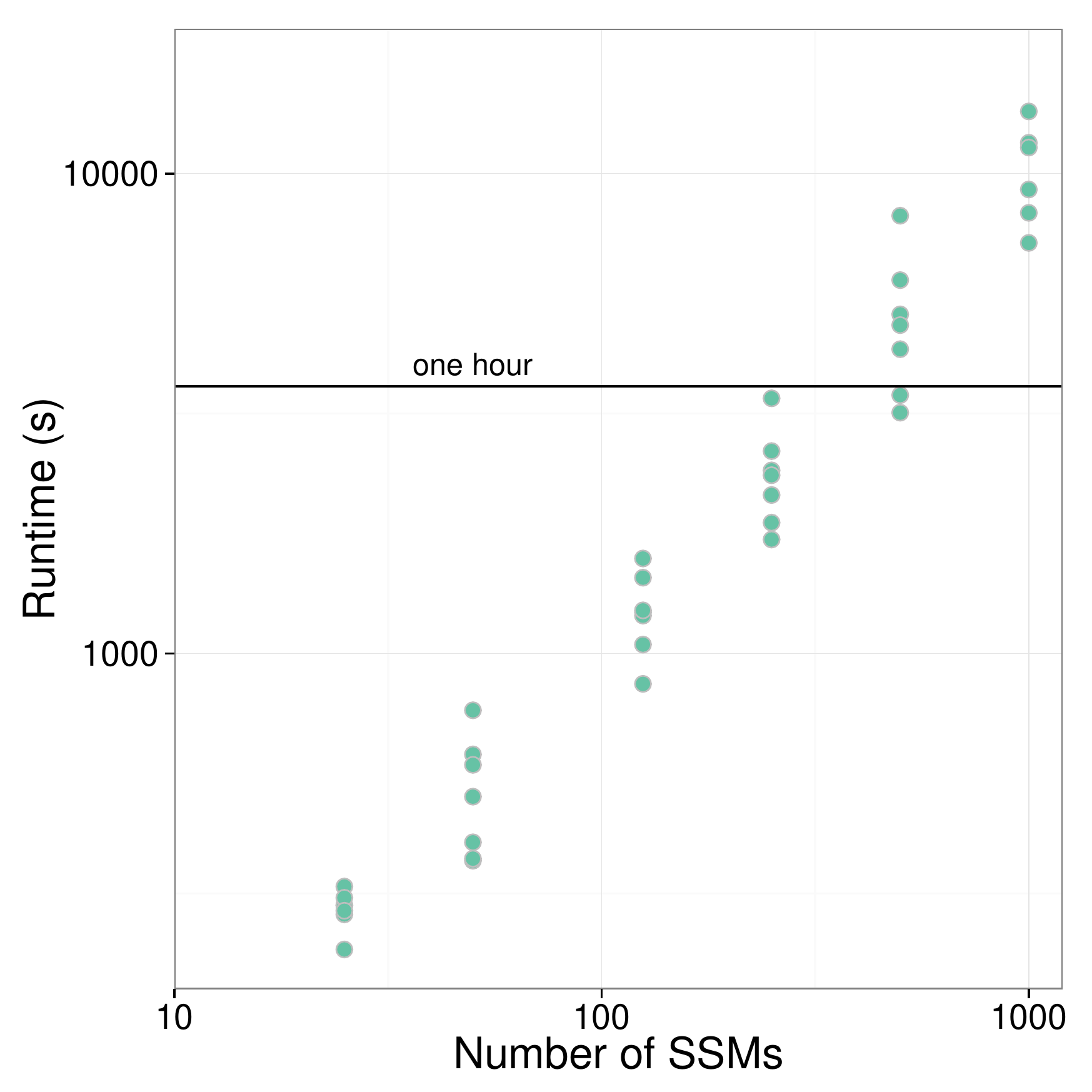}
\caption{{\bf PhyloWGS Runtime.}
The relationship between the number of SSMs in the simulated dataset with 5 subpopulations
and the runtime on a log10-log10 plot. Runtime is computed using a
single core of an Intel i7-4770K with 2500 MCMC iterations and
5000 inner Metropolis-Hastings iterations. Runtime can be greatly
decreased by parallelizing the sampling or by taking less samples;
however, the implications of these options have not been explored.}
\label{fig:runtime}
\end{figure}

An important question in subclonal analysis of tumor samples is
estimating how deep sequencing must be in order to recover the
subclonal structure.  To answer this question we applied PhyloWGS to simulated
read counts with known subclonal structure.  Our simulations looked at
a range of total population counts (3, 4, 5, 6), read depths (20, 30,
50, 70, 100, 200, 300) and number of SSMs per population (5, 10, 25,
50, 100, 200, 500, 1000). For each combination of population count,
read depth and SSMs per population we generated simulated tumor data
for which the subclonal population frequencies were consistent with both branching and linear phylogenies.  For each simulated SSM $k$ in subpopulation $u$, reference allele reads ($a_k$) were drawn as:
\begin{align*}
a_k \sim \text{Binomial}(d_k, 1-\phi_u + 0.5 \phi_u) \ ; \quad 
d_k \sim \text{Poisson}(r) \ ,
\end{align*}
where $\phi_u$ is the clonal frequency of population $u$ and $r$ is the simulated read depth.
A table of the $\phi$ values used for the simulations can be found as Table \ref{tbl:phis}.
\begin{table}[!t]
\caption{Table of subclonal lineage proportions used}
\begin{center}
\begin{tabular}{|l|l|}
Number of populations & $\phi$ values used (linear) \tabularnewline
\hline
3 & 0.44, 0.11\tabularnewline
4 &	0.56, 0.25, 0.06\tabularnewline
5 &	0.64, 0.36, 0.16, 0.04 \tabularnewline
6 &	0.71, 0.44, 0.25, 0.11, 0.03 \tabularnewline

\end{tabular}
\end{center}
\begin{flushleft}
\end{flushleft}
\label{tbl:phis}
\end{table}
First, we examined the time needed to complete sampling as a function of the number of SSMs (shown in Figure \ref{fig:runtime}).  
\highlight{In less than three hours on a single core of an Intel i7-4770K, on average, inference could be completed with up to 1000 SSMs (all timing data shown uses the simulated dataset with 5 subpopulations).}

\begin{figure}[!t]
\centering
\includegraphics[scale=.6]{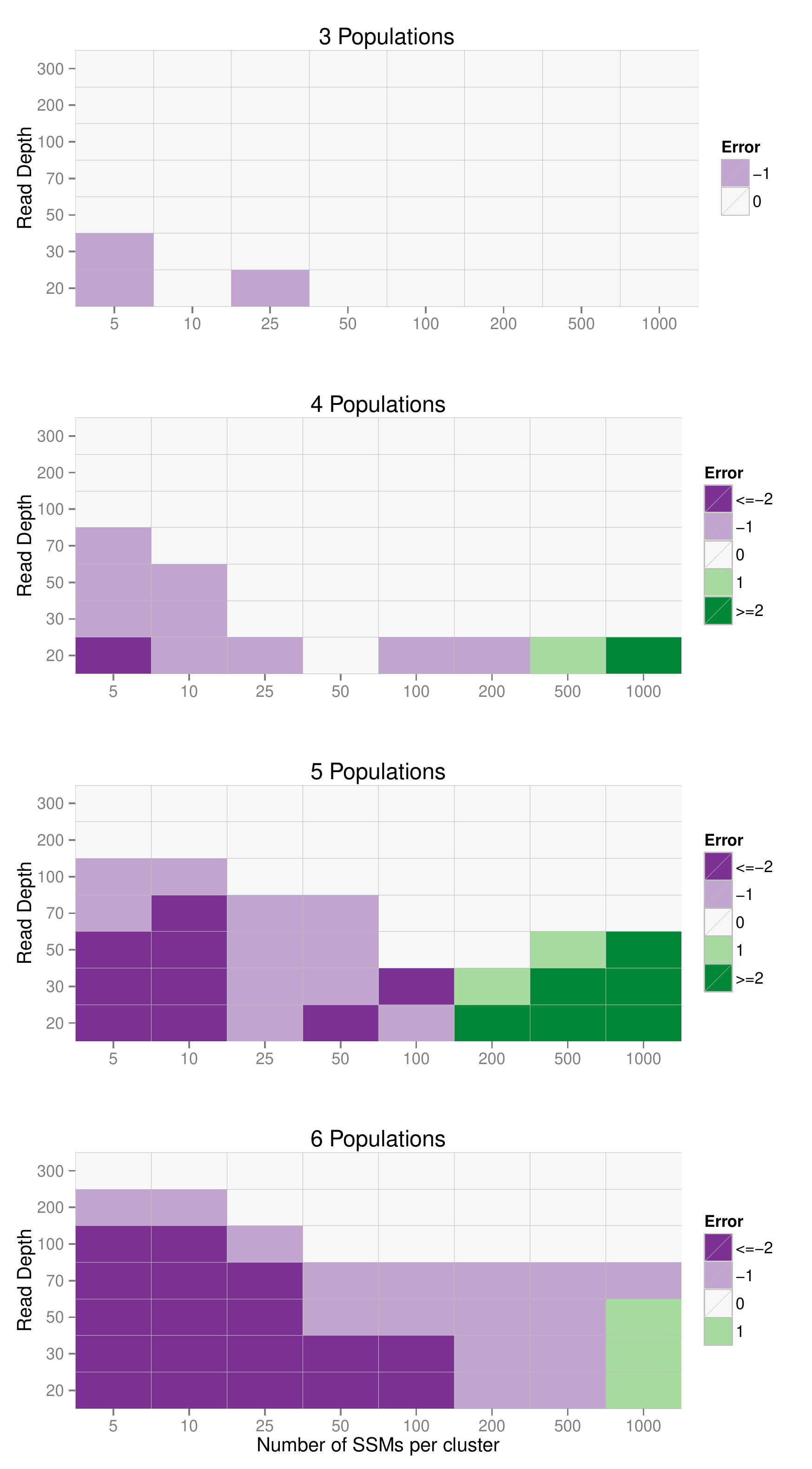}
\caption{{\bf Recovering true number of clusters.}
Each figure shows the relationship between the number of SSMs per cluster, the read depth and the ability of PhyloWGS to recover the true number of populations for 3 (A), 4(B), 5(C) and 6(C) population simulations. Error is calculated by subtracting the true number of subclonal lineages from the number found.}
\label{fig:cluster_error}
\end{figure}

To determine the number of subpopulations our algorithm found, we analyzed the sampled tree with the highest complete data likelihood and removed any subpopulations with zero assigned SSMs. We then compared the difference between the number of subpopulations used to generate the data and the number of subpopulations identified by our algorithm.  The results of this comparison for ambiguous phylogeny simulations are shown in Figure \ref{fig:cluster_error}.  
Several relationships between simulation parameters and the output of our model can be observed.  First, unsurprisingly, increasing the read depth and decreasing the number of subpopulations resulted in increased accuracy in the estimated number of subpopulations.   Second, for the ambiguous phylogeny simulations, there is a U-shaped relationship between accuracy and the number of SSMs characterizing each population, where accuracy first increases and then decreases as the number of SSMs increases. This decrease in accuracy with high numbers of SSMs is non-intuitive, as more SSMs provide more information with which to perform inference.  However, the Dirichlet process prior sometimes overestimates the number of source components \cite{millerharrison2013}.  While this overestimation has not been demonstrated for the tree-structured stick-breaking process used by PhyloWGS, the similarity between the processes makes it likely that this is the case.  While some of these errors can be eliminated by ad-hoc removal of clusters with a small number of SSMs, there is not yet a consistent approach to do this, so we leave the results untouched.  These results suggest that for three or four subpopulations, a read depth consistent with typical WGS experiments (20x--30x) is sufficient to identify the correct number of subpopulations, while experiments with 200x-300x are needed to resolve tumors with up to six subpopulations.

Another important measure of the performance of our algorithm is how
accurately the mapping from population to SSM is.  To evaluate this
accuracy in a systematic way that accounts for class-imbalance,
varying number of SSMs and differing number of clusters we examine the
Area Under the Precision-Recall Curve (AUPRC) between the known true
co-clustering matrix and the average co-clustering matrix from our
samples.
A co-clustering matrix $M$ is a binary matrix where $M_{ij} = 1$ if SSM $i$ and SSM $j$ are in the same cluster.
\highlight{The average co-clustering matrix is constructed by taking the
average of the co-clustering matrices of each sample in the
Markov chain after burn-in} and is an estimate of the
posterior mean co-clustering matrix of our model.
The average co-clustering matrix better predicts
the true co-clustering matrix than the co-clustering matrix computed
from the maximum data
likelihood tree. AUPRC was chosen over Area Under the
Receiver-Operator Curve (AUROC) as it is known to be more informative
in the presence of class-imbalance \cite{davis2006relationship} which
changes as the number of populations increase.

\begin{figure}[!t]
\centering
\includegraphics[scale=.6]{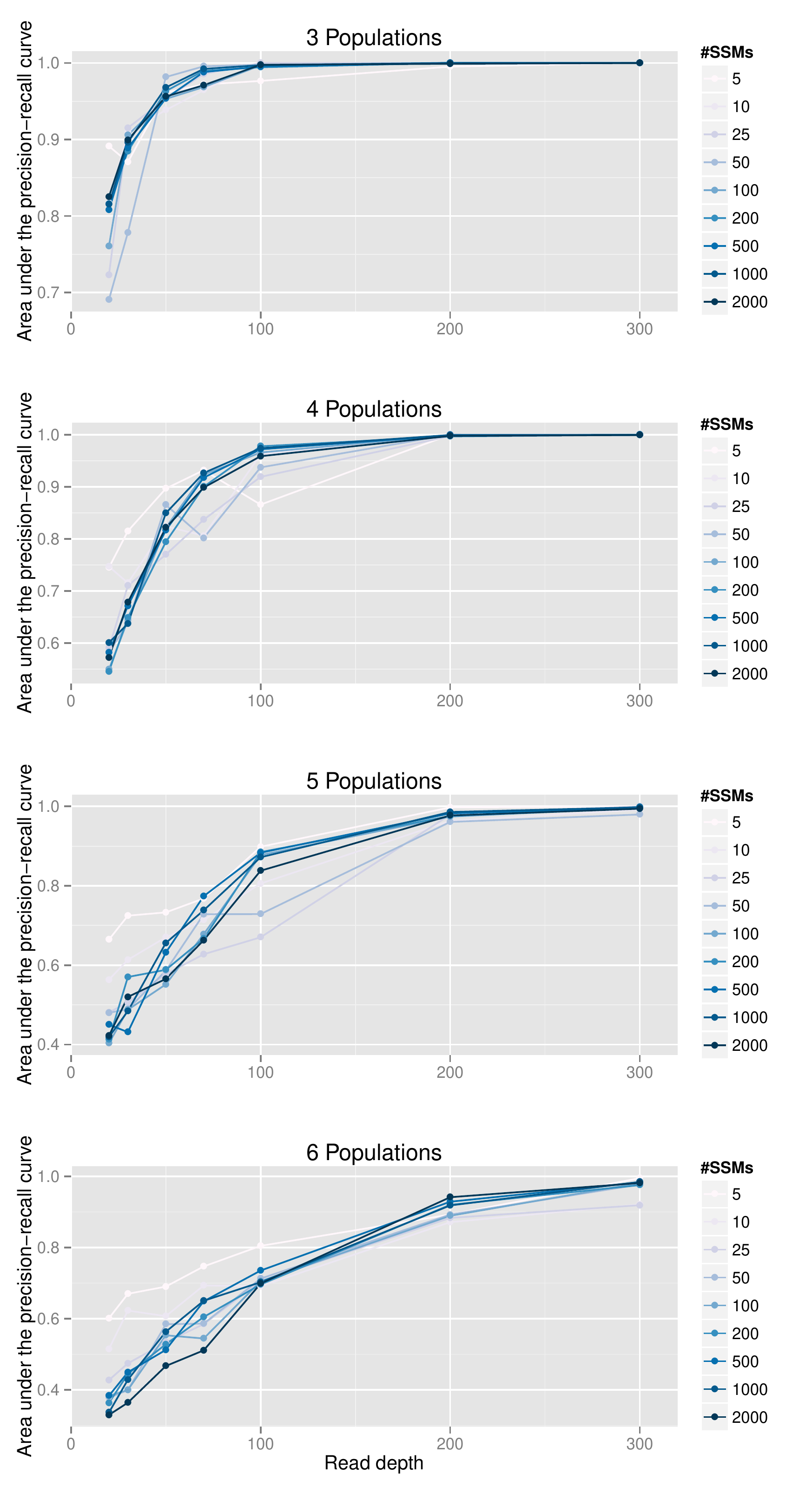}
\caption{{\bf Reconstruction accuracy}
Each figure shows the relationship between the read depth and the accuracy of the resulting clustering, measured as Area Under the Precision-Recall Curve (AUPR).  Plots for 3 (A), 4(B), 5(C) and 6(D) populations are shown with each line representing a different number of SSMs per cancerous population.}
\label{fig:aupr}
\end{figure}

\begin{figure}[!t]
\centering
\includegraphics[scale=.5]{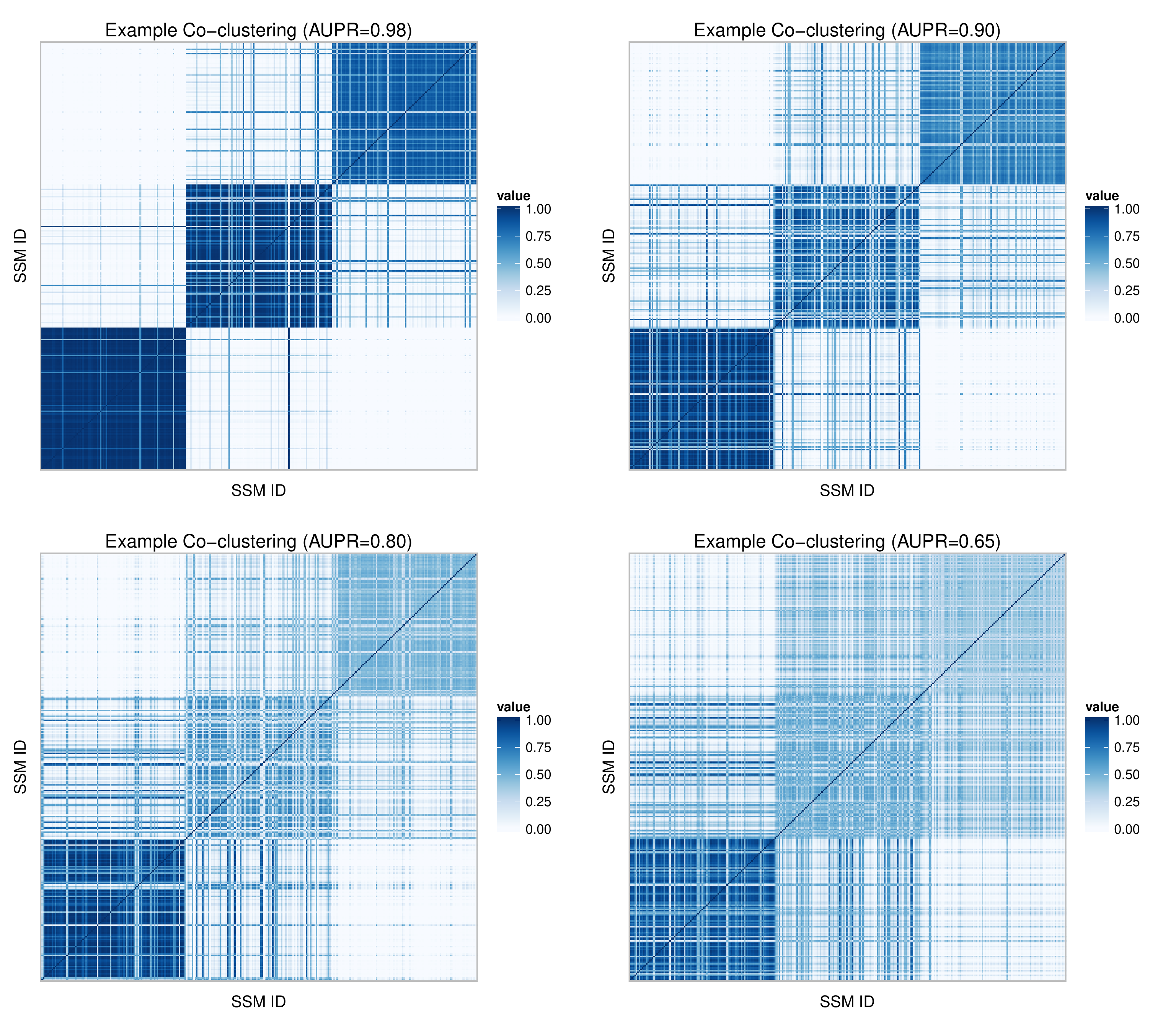}
\caption{{\bf Co-clustering examples}
Each figure shows mean co-clustering matrixes for simulations with 4
populations (3 cancerous), where the AUPR is 0.98 (A), 0.90 (B), 0.80
(C) and 0.65 (D). Rows and columns correspond to individual SSMs.
For visibility, the matrix has been randomly subsampled to 150 SSMs from the 600 SSMs used in the
simulation. Pixel color indicates co-clustering probability.}
\label{fig:cc_ex}
\end{figure}

In Figure \ref{fig:aupr} we plot the resulting AUPRC for our simulation experiments.  
As with inferring the number of populations, our method does better as the read depth increases and the number of populations decreases.  Unlike the last result, there is no clear relationship between the number of SSMs and the resulting AUPRC.  To provide qualitative guidance to users of the meaning of various AUPRC cutoffs, we show several examples of inferred co-clustering matrices with AUPRCs of 0.65, 0.8, 0.9 and 0.98 in Figure \ref{fig:cc_ex}.
  
%Similar plots showing AUROC instead of AUPRC can be found as Supplementary Figure 1.

\subsection{Simulations with CNV changes}
\begin{figure}[!t]
\centering
\includegraphics[scale=.55]{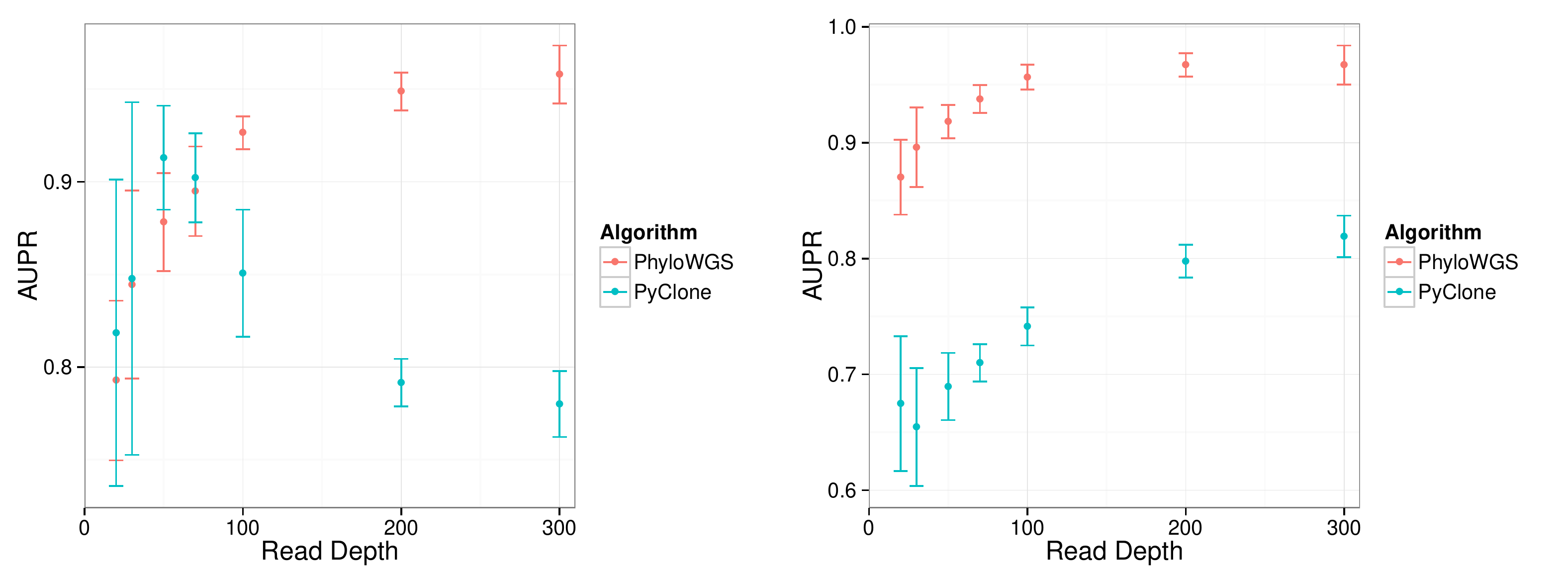}
\caption{{\bf Reconstruction accuracy}
Each figure shows the relationship between the read depth and the accuracy of the resulting clustering, measured as Area Under the Precision-Recall Curve (AUPR) for PhyloWGS and PyClone.  Plots are shown for subclonal additions (A) and deletions (B)}
\label{fig:subc}
\end{figure}

Next, we generated simulated data from a more complex genetic environment.  In these cases we simulated data from a tumor with 20\% normal tissue, a 40\% CNV-free subpopulation with 500 mutations and a descendent subpopulation with another 200 mutations but a substantial CNV affecting 50\% of the genome, either an amplification or a deletion.  We simulated data with read depth 20, 30, 50, 70, 100, 200 and 300 10 times for each read depth / alteration pair.  We then applied PhyloWGS and computed the AUPRC scores.  To demonstrate the importance of incorporating CNVs in phylogenetic reconstruction we compared the scores from PhyloWGS with those from PyClone \cite{pyclone}.  Performance for both methods can be seen in Figure \ref{fig:subc}.  
Using PhyloWGS results in superior clustering when compared to PyClone for both subclonal amplifications and deletions, with the exception of amplifications with low read-depths, where the performance distributions closely overlap.

\subsection{TCGA Benchmark}
\begin{figure}[!t]
\centering
\includegraphics[scale=.6]{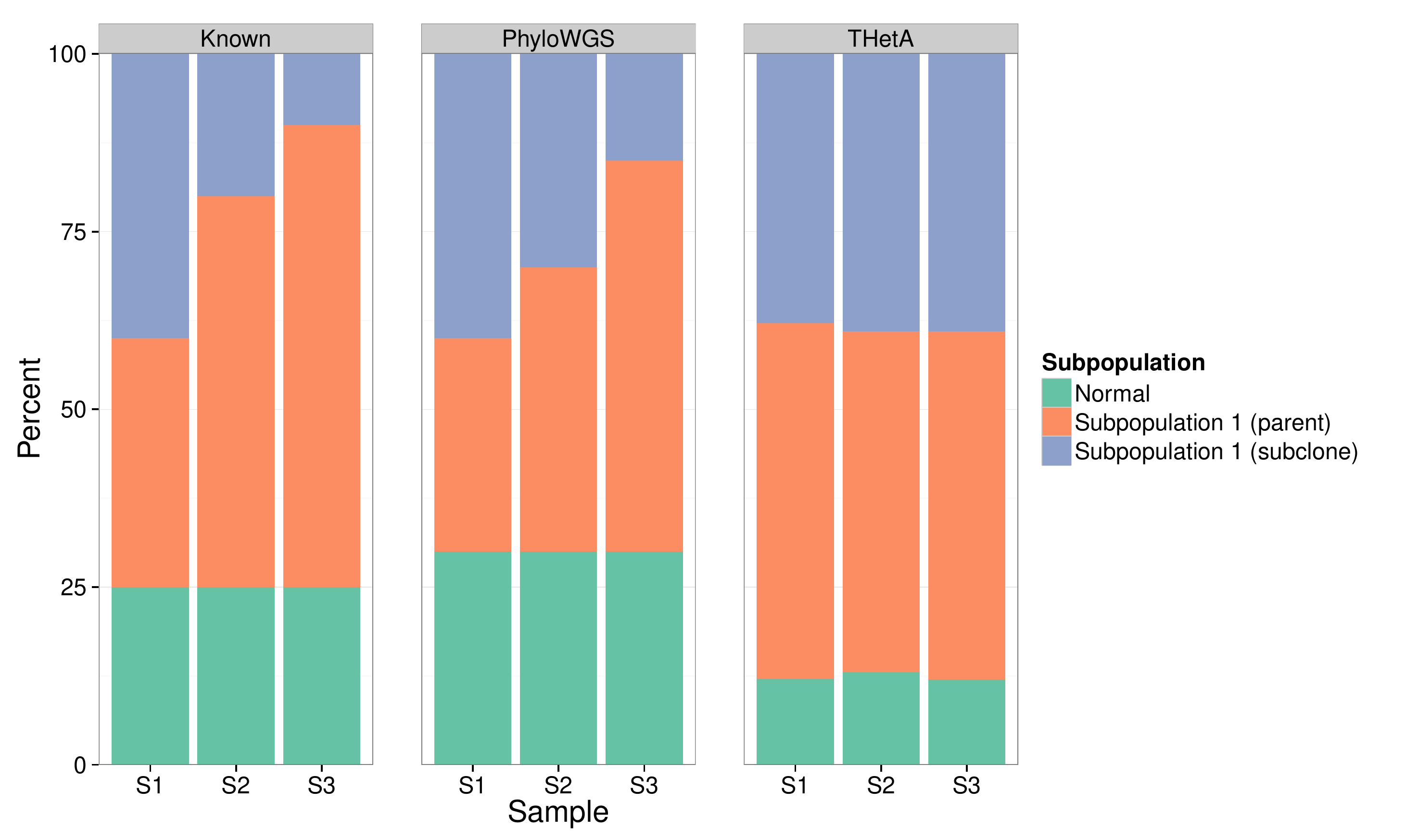}
\caption{{\bf True and inferred composition of TCGA benchmark samples.}
The figure shows the true (left), inferred by PhyloSub (center) and
inferred by THetA (right) composition of three TCGA benchmark
samples. Each bar represents a single sample.}
\label{fig:newtcga}
\end{figure}

Next, we apply PhyloWGS to the TCGA variant-calling benchmark 4 dataset \cite{tcga}.  The samples we examined consist of a normal population, a cancerous cell-line population HCC 1143 and a spiked-in subclonal descendant of the cancerous population in various proportions with 30x coverage. Starting with the publicly available BAM files we identified locations of possible structural variation using BIC-seq \cite{bicseq} with default parameters, except for the bandwidth parameter, which was set to 1000.  We changed the bandwidth parameter because we found the default value of 100 resulted in overly noisy segmentations and highly variable normalized read counts.  To identify SSMs and the number of variant and reference reads for each SSM, we reverted the BAM files into unaligned reads using Picard 1.90 \cite{picard}. Reads for each sample were then realigned using BWA 0.6.2 \cite{bwa} and collapsed using Picard. Aligned reads of a cancerous sample and its matched normal were analyzed by two somatic calling tools: MuTect 1.1.4 \cite{mutect} and Strelka 1.0.7 \cite{strelka}. A set of high confidence mutations were extracted by taking an intersection of the calls made by MuTect and Strelka. Previous verification on other tumor/normal pairs showed that this approach achieved $>90\%$ precision (data not shown).
We first ran THetA \cite{theta} using the output of BIC-seq with the
aim of using THetA's output to provide us with the CNV information
that PhyloWGS requires (see \textrm{Methods} section).  However,
despite the fact that the subclonal population varied from 40\% to
10\%, THetA returned nearly identical composition inferences for all
the samples (see Figure \ref{fig:newtcga}). Because of this, we
decided that we could not rely on THetA's copy number calls, we instead simply removed all SSMs in a location where BIC-seq identified possible structural variation.  This eliminated most of the SSMs identified, leaving only 62 SSMs from the original 4,344.  Despite this small number of SSMs our algorithm was still able to identify the correct number of populations and captured the changing composition of the samples. Also, the inferred SSM content of each cluster was identical in the three separate runs.

\subsection{Chronic Lymphocytic Leukemia}
\begin{figure}[!t]
\centering
\includegraphics[scale=1]{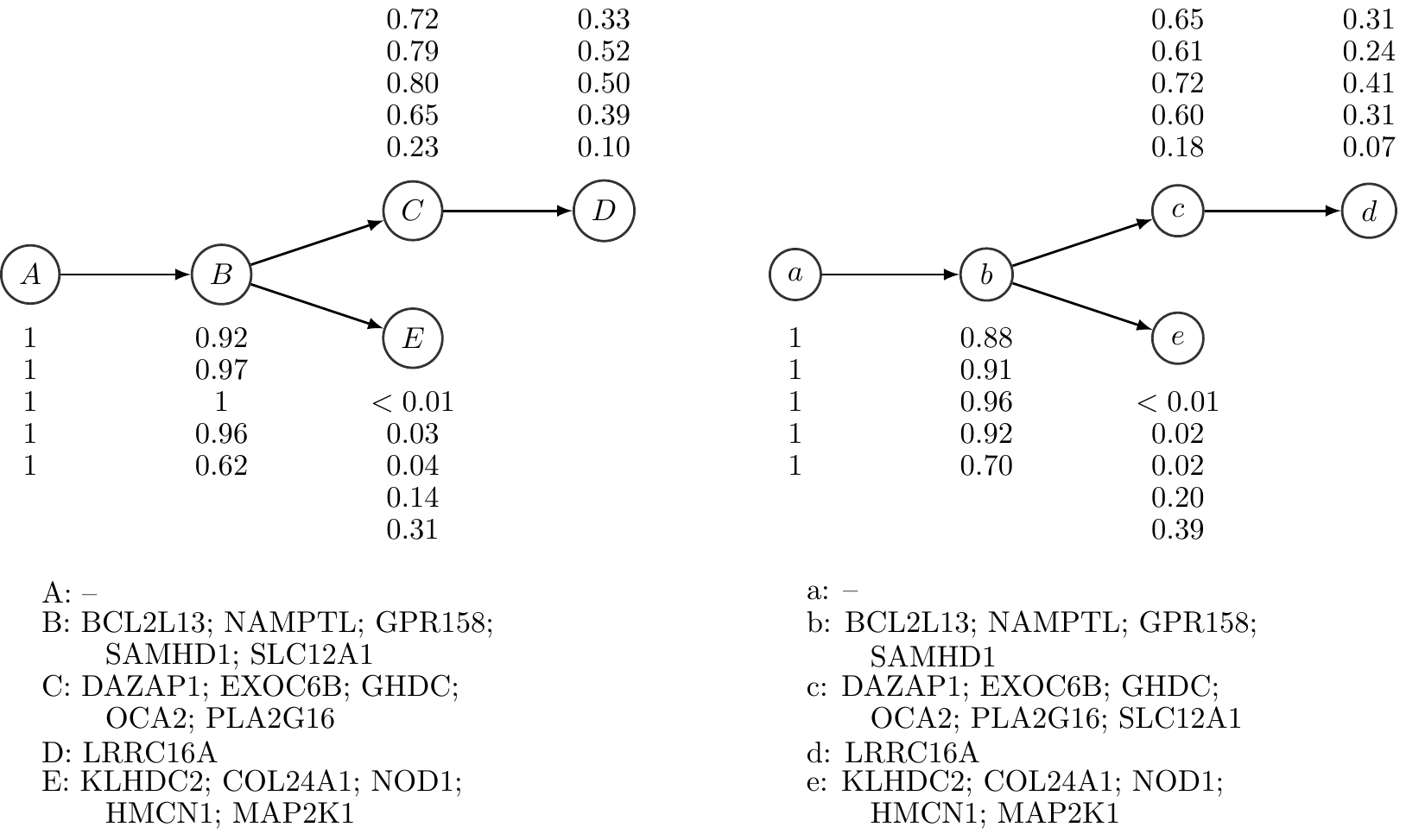}
\caption{{\bf Expert-generated and inferred phylogenies from CLL patient CLL077.}
Left: The expert-generated phylogeny based on targeted deep sequencing data.
Right: The phylogeny inferred by PhyloWGS on allele frequencies of the
same SSMs found using WGS. The subclonal lineage population
frequencies for the five samples and the SSM assignments of lineages are also shown in the figure.}
\label{fig:cll077_wgs}
\end{figure}

Next, we applied PhyloWGS to a data from patient CLL077 extracted from
Supplementary Table 7 from a paper describing a Chronic Lymphocytic Leukemia (CLL) dataset\cite{Schuh12}. For this patient, five tumor samples were collected over the course of treatment. We note that our method does not assume or use any temporal relationships in multiple sample data and could equally be applied to multiple samples collected simultaneously.  
We have previously reported experiments using the targeted
resequencing data with average read depth of 100,000x at 17 identified
SSMs \cite{phylosub}, we now instead use the data from WGS for that
same set of mutations, with average read depth of 40x. By examining
the number of reference and variant alleles it was clear that the
mutation in gene SAMHD1 was at a location that was homozygous in the
cancerous subpopulation it was part of.    This is because the
proportion of variant reads was far above 50\% (the expected variant
allele proportion for a heterozygous SSM present in every cell of the
sample).  We simulated the data that a CNA algorithm would find by
assuming that the copy number at that location was one in a CNA-defined subpopulation and that the proportion of cells in that population was the same as implied by halving the proportion of variant alleles.  After running PhyloWGS on this data, we compare the maximum data likelihood tree with the expert-generated tree found using a semi-manual method and targeted resequencing data in Figure \ref{fig:cll077_wgs}.  The two trees are nearly identical with the exception of assigning a single SSM to a child of the subpopulation where it is found in the expert tree. 
%In Additional File 2, we show the top 50 sampled trees, ranked based on their posterior probabilities.

\subsection{Breast Tumor}
\begin{figure}[!t]
\centering
\includegraphics[scale=1]{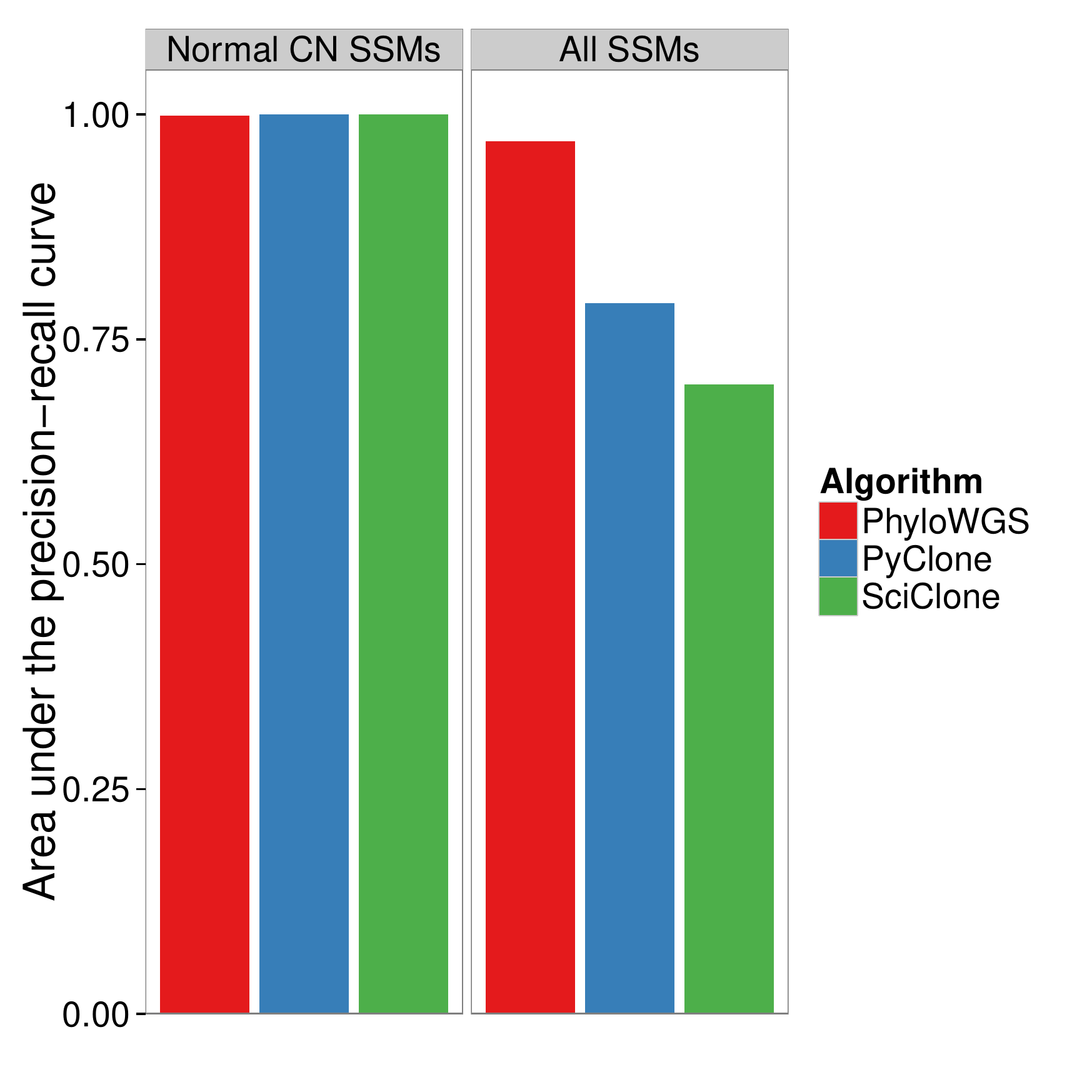}
\caption{{\bf Subclonal reconstruction algorithms applied to breast tumor PD4120}
Left: Area Under the Precision Recall curve (AUPR) for PhyloWGS, PyClone and SciClone when looking at SSMs in areas of normal copy number.
Right: AUPR for PhyloWGS, PyClone and SciClone when looking at SSMs in altered and normal copy number.}
\label{fig:pd4120}
\end{figure}

We analyzed data from WGS at 288x coverage from tumor PD4120a, first
reported in \cite{nik12} and re-analyzed in \cite{theta}.  We confined
our analysis to SSMs in genomic regions where THetA and the original analysis
agreed on the copy number status of the genome (chr 3,4q,5,10,13,16q,17,19 and 20).  
These regions contains a total of 26,029 SSMs, of which 4,739 were in regions affected by clonal copy number changes and 2,171 were in regions affected by subclonal copy number changes.  
We then ran PhyloWGS, PyClone and SciClone on SSMs in regions of normal copy number and on
SSMs in both altered and normal copy number.
PyClone uses a non-phylogenic correction for copy number alterations and SciClone performs no correction.
Based on the semi-manual clustering
from \cite{nik12}, we identified those mutations assigned to
clusters D,C and B with high probability to use as our gold standard
for clustering.  We then compared AUPR for all three algorithms on the
two datasets (see Figure \ref{fig:pd4120}).  All three algorithms have
very similar performance when only looking at SSMs in normal regions (Fig. \ref{fig:pd4120} left panel).
PhyloWGS continues to have very high performance when SSMs in regions
of 
copy number alterations are included, while both PyClone and SciClone have
much worse performance than PhyloWGS (Fig. \ref{fig:pd4120} right panel).

%%%%%%%%%%%%%%%%%%%%%%%%%%%%%%%%%%%%%%%%%%%%%%%%%%%%%%%%%%%%%%%%
%%%%%%%%%%%%%%%%%%%%%%%%%%%%%%%%%%%%%%%%%%%%%%%%%%%%%%%%%%%%%%

\section{Discussion}
Our work makes two important contributions to the burgeoning field of
subclonal reconstruction.
First, we provide the first automated method that integrates SSM and
CNV data in the reconstruction of tumor phylogenies.
This is an important innovation, previous methods either ignore the
impact that CNVs have on SSM allele frequencies \cite{trap,purbayes},
or assume that the CNVs affect all (and only) the cells that contain
the SSM \cite{pyclone, phylosub, expands}.
These assumptions can lead to incorrect inferences about the population
frequency of SSMs because how a CNV affects the allele frequency of
an SSM depends on its phylogenic relationship with the SSM.
Many of the insights about how to integrate SSM and CNV
data appear in \cite{nik12}; our work here extends and formalizes
these seminal observations while also providing an automated method
for phylogenic reconstruction.
A further advantage of combining SSMs and CNVs in the phylogenic
reconstruction is that CNVs overlapping the SSM locus can provide
further constraints on the tree-structure than are provided by SSM
frequency alone, and we described one case where it is possible to
unambiguously infer branching when an amplification of a
SSM-containing locus does not lead to a large increase in the SSM
allele frequency.
Second, we show that given typical WGS read depths,
SSM-based methods are able to accurately reconstruct tumor phylogenies
and detect and assign SSMs for at least six subpopulations.
Previously it wasn't clear to what extent this
reconstruction is possible; and no automated reconstructions with more than two
cancerous subpopulations based on WGS data had been described.
\highlight{Furthermore, we demonstrate the importance of phylogenetic correction of VAFs of SSMs
that occur in loci affected by copy number changes when performing
subclonal reconstruction. Specifically, we presented results on a
breast cancer benchmkar where methods that do not use PhyloWGS's
phylogenic correction perform much worse at recovering subpopulations.}
Finally, we report examples of accurate subclonal reconstruction
for cancer populations with highly reordered chromosomes solely on the
basis of SSM frequencies in the regions of normal copy
number. On these same data, a state-of-the-art CNV-based method failed
to perform the reconstruction. 

The current version of PhyloWGS relies on preprocessing the sequencing
data with a CNV-based method for subclonal reconstruction.
This is because it assumes that initial population frequency $\phi_{i}$ and
copy number data $C_i$ are already available for the CNVs;
furthermore, for amplifications, $C_i > 2$, it requires $C_i$ to be
separated into the relative number of each of the two copies, i.e., $\{C_i^m, C_i^p\}$.
It does not, however, require the SSMs to be phased; in other words,
it does not need to know whether the SSMs occurred on the maternal or
paternal copy of the chromosome.
New CNV-based methods provide $\{C_i^m,
C_i^p\}$\cite{titan}; our work anticipates further developments in
subclonal CNV callers.
If used with a method that cannot decompose copy number changes into
changes in the maternal and paternal loci, PhyloWGS can be restricted to regions of copy number
loss (i.e. $C_i < 2$), where there is only one possible
breakdown. Note that this decomposition is only necessary when an SSM precedes a CNV on the
same branch of a phylogeny.

We also note that PhyloWGS does not require the CNV-based preprocessing to
be able to detect all of the subclonal populations, and we have shown
that PhyloWGS can detect additional populations either defined completely by
SSMs or that were not detected by CNV-based methods.
This is particularly important because recent CNV-based methods are limited
to a maximum of two cancerous populations and those that
can detect $>1$ cancerous subpopulation do so by relying on a strong
parsimony assumption.
If invalid, this assumption can lead to large errors in subclonal reconstruction
because it selects branching phylogenies
over chain phylogenies that are equally well-supported by the data.

Indeed our results suggest an alternative strategy for combining SSMs
and CNVs in subclonal reconstruction.
Regions unaffected by CNVs can be relatively easily detected using
methods such as BIC-seq \cite{bicseq}.
Even in highly rearranged cancer genomes, there are often
non-negligible regions of normal copy number and we have
shown that we can achieve reasonably accurate subclonal
reconstructions using the limited number of SSMs in regions of normal copy
number in the TCGA benchmark.
\hide{As such, we propose that SSM-based subclonal reconstruction should be
performed first on SSMs in regions of normal copy number.
This initial reconstruction can then guide the assignment of CNVs to
subpopulations: the population frequencies derived from the initial
SSM-based reconstruction simplify the inference of CNVs by removing
one of the unknowns normally present in CNV-based reconstruction.
Finally, the SSMs in CNV-affected regions can then be assigned to
subpopulations.
A major advantage of this approach is that it avoids the problematic strong parsimony
assumption that guides current CNV-based methods.
However, this approach would miss subclonal lineages defined purely by CNVs.}

In the final stages of preparing this manuscript, a new method, cloneHD \cite{cloneHD} was published. 
Like PhyloWGS, this method combines both SSMs and CNVs in subclonal reconstruction and does so using WGS data from single and multiple samples. 
However, unlike PhyloWGS, cloneHD does not explicitly perform phylogenic reconstruction, so it is unable to fully account for the phylogenic relationship among SSMs and CNVs when analyzing SSM allele frequency. 
As such, it is not clear to us that it can correctly solve the subclonal reconstruction problem posed in Figure 1. 
The cloneHD manuscript also does not extend the limits of WGS-based subclonal reconstruction as none of the examples reconstruct more than two cancerous subpopulations from a single sample. 
Finally, cloneHD appears to rely on the strong parsimony assumption in order to assess subclonal genotypes, and only reports a single reconstruction, obscuring the uncertainty involved. However, cloneHD does appear to be an interesting and powerful method and we hope that future work can compare the merits and drawbacks of these alternate approaches to subclonal reconstruction.

%%%%%%%%%%%%%%%%%%%%%%%%%%%%%%%%%%%%%%%%%%%%%%%%%%%%%%%%%%%%%%
%%%%%%%%%%%%%%%%%%%%%% CONCLUSIONS %%%%%%%%%%%%%%%%%%%%%%%%%%%%%%
\section{Conclusions}

We have presented a new method, PhyloWGS, that reconstructs tumor phylogenies and
characterizes the subclonal populations present in a tumor sample using
both SSMs and CNVs.
Our method takes as input measures of allelic frequency of SSMs, as
well as estimates of population frequencies and copy number for each
CNV. PhyloWGS groups the SSMs and CNVs into subpopulations, and
estimates the population frequencies and the phylogenic relationship of
these subpopulations.
PhyloWGS is based on a generative probabilistic model of allele
frequencies that incorporates a nonparametric Bayesian prior over trees.
The output of PhyloWGS consists of samples from this
distribution generated through Markov Chain Monte
Carlo and we report the tree that maximizes the likelihood
of the data found during the sampling run, if a single point
estimate is required.
However, unlike our previous PhyloSub method \cite{phylosub}, PhyloWGS includes CNVs in its subclonal
reconstruction and, in doing so, can correctly account for the effect
of CNVs on the VAF of overlapping SSMs.
PhyloWGS also runs more than 50 times faster than PhyloSub, making it feasible to
apply it to the thousands of SSMs that are found through WGS. 

We have applied PhyloWGS to real and simulated data from WGS of
tumor samples to demonstrate that subclonal populations can be
reliably reconstructed based solely on SSMs from medium depth
sequencing (30x-50x). We have also used PhyloWGS to correctly solve a
simulated subclonal reconstruction problem that neither an SSM-based nor
CNV-based method can solve alone; and to reconstruct the phylogeny and
subclonal composition of a highly rearranged sample for which a CNV-based method fails. 
We also demonstrate that when applied to WGS of time-series samples from a chronic lymphocytic
leukemia patient, PhyloWGS recovers the same tumor phylogeny previously
reconstructed by applying PhyloSub and a semi-manual method to data
from deep targeted resequencing. 
\highlight{Finally, we demonstrate state-of-the-art performance in subclonal reconstruction on a breast tumor sample, highlighting the advantages of performoing phylogenetic CNV correction.}
Our work thus greatly expands the range of tumor samples for which
subclonal reconstruction is possible, enabling widespread use of
automated subclonal reconstruction for medium-depth WGS sequencing experiments.

%%%%%%%%%%%%%%%%%%%%%%%%%%%%%%%%%%%%%%%%%%%%%%%%%%%%%%%%%%%%%%%%
%%%%%%%%%%%%%%%%%%%%%%%%%%%%%%%%%%%%%%%%%%%%%%%%%%%%%%%%%%%%%%

%%%%%%%%%%%%%%%%%%%%%%%%%%%%%%%%%%%%%%%%%%%%%%%%%%%%%%%%%%%%%%
%%%%%%%%%%%%%%%%%%%%%% METHODS %%%%%%%%%%%%%%%%%%%%%%%%%%%%%%
\section{Methods}
\subsection{PhyloSub model}
Our probabilistic model for read count data is based on PhyloSub \cite{phylosub}. For each SSM that is detected by high-throughput sequencing methods, cells containing the SSM are referred to as the variant population and those without the variant as the reference population. Let $a_i$ and $b_i$ denote the number of reads matching the reference allele $\textrm{A}$ and the variant allele $\textrm{B}$ respectively at position $i$, and let $d_i = a_i+b_i$. Let $\mu_i^r$ denote the probability of sampling a reference allele from the reference population. This value depends on the error rate of the sequencer. Let $\mu_i^v$ denote the probability of sampling a reference allele from the variant population which is set to 0.5 if there are no CNVs -- in other words, the SSM is assumed to only affect one of the two chromosomal locations. Let $\tilde{\phi}_i$ denote the fraction of cells from the variant population, i.e., the SSM population frequency at position $i$, and $1-\tilde{\phi}_i$ denote the fraction of cells from the reference population at position $i$.  Let $\text{DP}(\alpha,H)$ denote the Dirichlet process (DP) prior with base distribution $H$ and concentration parameter $\alpha$. Samples from the DP are used to generate the SSM population frequencies $\{\tilde{\phi_i}\}$.  The observation model for allelic counts has the following generative process:
\begin{equation}\label{eqn:pm}
\begin{aligned}
& \mathcal{G}  \sim  \text{DP}(\alpha,H) \ ; \quad
\tilde{\phi}_{i}  \sim \mathcal{G} \ ; \quad
& a_i \mid d_i,\tilde{\phi}_i,\mu^r_i,\mu^{v}_i  \sim \text{Binomial}(d_i,(1-\tilde{\phi}_i)\mu^r_i+\tilde{\phi}_i \mu_i^{v}) \ .
\end{aligned}
\end{equation}
The posterior distribution of $\tilde{\phi}_i$ is $p(\tilde{\phi}_i \mid a_i,d_i,\mu_i^r,\mu_i^{v}) \propto p(a_i \mid d_i,\tilde{\phi}_i,\mu_i^r,\mu_i^{v}) p(\tilde{\phi}_i)$.

The Dirichlet process prior $\text{DP}(\alpha,H)$ in the observation model of allelic counts \eq{eqn:pm} is useful to infer groups of mutations that occur at the same SSM population frequency \cite{Shah12,pyclone}. Furthermore, being a nonparametric prior, it allows us to avoid the problem of selecting the number of groups of mutations \emph{a priori}. However, it cannot be used to model the clonal evolutionary structure which takes the form of a rooted tree. In order to model this, we use the tree-structured stick-breaking process prior \cite{AdamsGJ10} denoted by $\text{TSSB}(\alpha,\gamma,H)$. The parameters $\alpha$  and $\gamma$ influence the height and width of the tree respectively and are similar to the concentration parameter in the Dirichlet process prior. Let $\{\phi_k\}_{k=1}^K$ denote the set of unique frequencies in the set $\{\tilde{\phi}_i\}_{i=1}^N$, where $K$ is the number of subclones or nodes in the tree. In other words, multiple elements in the set $\{\tilde{\phi}_i\}_{i=1}^N$ will take on the same value from the set $\{\phi_k\}_{k=1}^K$ of unique frequencies. The prior/base distribution $H$ of the SSM population frequencies is the uniform distribution $\text{Uniform}(0,1)$ for the root node and $\text{Uniform}(0,\phi_{\text{par}(v)}-\sum_{w \in \mathcal{S}(v)} \phi_w)$ for any other node $v$ in the tree, where $\text{par}(v)$ denotes the parent node of $v$ and $\mathcal{S}(v)$ is the set of siblings of $v$. This ensures that the clonal evolutionary constraints (discussed below) are satisfied when adding a new node in the tree. Given this model and a set of $N$ observations $\{(a_i,d_i,\mu_i^r,\mu_i^v)\}_{i=1}^N$, the tree structure and the SSM population frequencies $\{\tilde{\phi}_i\}$ are inferred using Markov Chain Monte Carlo (MCMC) sampling  (see PhyloSub \cite{phylosub} for further details). 

Given the current state of the tree structure, we sample SSM population frequencies in such a way that the SSM population frequency $\phi_v$ of every non-leaf node $v$ in the tree is greater than or equal to the sum of the SSM population frequencies of its children. To enforce this constraint, we introduce a set of auxiliary variables $\{\eta_v\}$, one for each node, that satisfy $\sum_v \eta_v =1$, and rewrite the observation model for allelic counts \eq{eqn:pm} explicitly in terms of these variables resulting in the following posterior distribution:
\begin{equation}
\label{eqn:posterior1}
p(\tilde{\eta}_i \mid a_i,d_i,\mu_i^r,\mu_i^{v}) \propto p(a_i \mid d_i,G_i = g,\tilde{\eta_i},\mu_i^r,\mu_i^{v}) p(\tilde{\eta}_i) \ ,
\end{equation}
where we have used $\{\tilde{\eta}_i\}$ to denote the auxiliary variables for each SSM. The prior/base distribution of the auxiliary variables is defined such that it is 1 for the root node and $\text{Uniform}(0,\eta_{\text{par}(v)})$ for any other node $v$ in the tree. When a new node $w$ is added to the tree, we sample $\eta_w \sim  \text{Uniform}(0,\eta_{\text{par}(w)})$ and update $\eta_{\text{par}(w)} \gets \eta_{\text{par}(w)} - \eta_w$. This ensures that $\sum_v \eta_v =1$. This change is crucial as it allows us to design a Markov chain that converges to the stationary distribution of  $\{\eta_v\}$. The SSM population frequency for any node $v$ can then be computed via  $\phi_v= \eta_v + \sum_{w \in \mathcal{D}(v)} \eta_w = \eta_v + \sum_{w \in \mathcal{C}(v)} \phi_w$, where $\mathcal{D}(v)$ and $\mathcal{C}(v)$ are the sets of all descendants and children of node $v$ respectively. This construction ensures that the SSM population frequencies of mutations appearing at the parent node is greater than or equal to the sum of the frequencies of all its children.  We use the Metropolis-Hastings algorithm \cite{hastings1970monte} to sample from the posterior distribution of the auxiliary variables $\{\tilde{\eta_i}\}$ \eq{eqn:posterior1} and derive the SSM population frequencies from these samples by selecting the sampled set of population frequencies with the highest likelihood. We use an asymmetric Dirichlet distribution as the proposal distribution. 

\subsection{Integrating CNV data into PhyloSub}
\label{PhyloWGS}
The focus of our new method, PhyloWGS is integrating SSM frequencies with existing copy number variation (CNV) based subclonal reconstructions.  As mentioned above, our algorithm takes as input a set of SSMs along with their allele frequencies, expressed for each SSM $i$, as the number of reads at the locus supporting either the SSM ($b_i$) or the reference allele ($a_i$). We also allow our algorithm to take a set of inferred copy number changes, where for each change $j$, the input provides the new copy number $C_{j}$ as well as the proportion of the population with the change $\tilde{\phi}_{j}$. In some cases, we also require the breakdown of $C_{j}$ into the new number of maternal ($C_{j}^m$) and paternal ($C_{j}^p$) copies of the locus (see below for details). If this breakdown is not available, we can restrict our attention to CNVs for which $C_{j} < 2$ because in these cases, there is only one possible breakdown. Also, in the absence of paternal/maternal breakdown, we should still be able to, in theory, assign SSMs with overlapping CNVs with $C_{j} > 2$ to specific populations once the phylogeny and subclonal populations have been defined using SSMs and CNVs in regions of $C_{j} \leq 2$.   

Below, we describe the rules, based on the infinite sites assumption, that we use to determine the relationship between the population frequency of an SSM $\tilde{\phi}_{i}$ and its observed variant allele frequency $(b_i/d_i)$. When the SSM does not overlap a region that has a predicted CNV in any cell in the tumor population, then the predicted allele frequency is simply half of the modeled population frequency. We also describe the method by which we transform each CNV $j$ into a pseudo-SSM to be included in the phylogeny.

\subsubsection{If CNVs do not overlap with any SSM} 
If a CNV occurs in a region without any SSMs, we generate a `pseudo-SSM' for the CNV $j$ which is represented in the model as a heterozygous, binary somatic mutation with a read depth that reflects the uncertainty in the provided population frequency $\tilde{\phi}_{j}$ for the CNV. Specifically, we generate reference and variant read counts, $a_j$ and $b_j$, respectively, such that the allelic frequency $b_j / (a_j + b_j)$ is approximately equal to $\phi_{j} / 2$ and the total number of reads $a_j + b_j$ is selected based on the evidence supporting the CNV. Generating this pseudo-SSM allows the CNV to be treated like any other SSM in the phylogeny model.

\subsubsection{If CNVs overlap with SSMs}
If a structural variant occurs in a region with an SSM $i$, this complicates the relationship between the proportion of cells that contain the SSM and the expected number of reads because cells with the CNV will have more (or fewer) than two copies of the locus where the SSM lies. Assuming equal sampling of these regions, the expected proportion of reads without the mutation ($\zeta_i$) is always: ${N_i^r}/{(N_i^r+N_i^v)}$ where $N_i^r$ is the number of copies of the locus that have the reference allele and $N_i^v$ is the number of copies of the locus with the variant allele. To account for sequencing error we define $\epsilon$ as the probability of reading the reference allele when the locus contains the variant allele and vice-versa.  The expected proportion of reads containing the reference allele is then:
\begin{align*}
\zeta_i = \frac{N_i^r(1-\epsilon) + N_i^v \epsilon}{N_i^r + N_i^v}
\end{align*} 
Looking at a tumor sample with multiple populations and without structural variations, if each population $u$ is present with proportion $\eta_u$ and where $s_i^u$ is $1$ if population $u$ contains the SSM $i$ and $0$ otherwise, then $N_i^r = 2 * \sum_u \eta_u (1-s_i^u) +\sum_u \eta_u s_i^u$ and $N_i^v = \sum_u \eta_u s_i^u$.  This is equivalent to an algorithm that looks at each population and performs the following update.  If the population $u$ contains the SSM $i$ then
\begin{align*}
N_i^r &\leftarrow N_i^r + \eta_u  \ , \\
N_i^v &\leftarrow N_i^v + \eta_u \ .
\end{align*}
If the population does not contain the SSM then:
\begin{align*}
N_i^r &\leftarrow N_i^r + 2\eta_u \ , \\
N_i^v &\leftarrow N_i^v + 0 \ .
\end{align*}
To take into account CNVs requires a more complex procedure.  For each population, for each SSM, the number of reference and variant alleles depends on the copy number of the locus $C_i$ and, potentially, number of maternal ($C_i^m$) and paternal ($C_i^p$) copies of the locus \emph{as well as} the evolutionary relationship between the SSM and the CNV.  The infinite sites assumption does not apply for CNVs, adding a further level of complexity because multiple CNVs at the same locus are possible. For each population, the CNV that affects its contribution to the number of reference and variant genomes can be found by ascending the evolutionary tree towards the root.  The first CNV found in this ascent is the CNV relevant for the population.  If no CNV is found than the population is not affected by a CNV.  For each population there are five possible situations:
\begin{enumerate}[(i)]
\item The population does not contain the SSM and is not affected by a CNV
\item The population does not contain the SSM but is affected by a CNV
\item The population contains the SSM but is not affected by a CNV
\item The population contains the SSM and is affected by a CNV, and the SSM occurred after the CNV
\item The population contains the SSM and is affected by a CNV, and the CNV occurred after the SSM
\end{enumerate}

If a population does not contain the SSM, then even if a CN change has occurred (cases i and ii), the update rule is:
\begin{align*}
N_i^r &\leftarrow N_i^r + \eta_u C_i \ , \\
N_i^v &\leftarrow N_i^v +0 \ .
\end{align*}
If a population contains the SSM and the SSM occurred after a CN change (or there was no CN change) (cases iii and iv) then there is a single copy of the mutated genome and the remainder are reference, so the the update rule is:
\begin{align*}
N_i^r &\leftarrow N_i^r + \eta_u \times \max(0,C_i-1) \ , \\
N_i^v &\leftarrow N_i^v + \eta_u \ .
\end{align*}
If a population contains the SSM and the SSM occurred before the CN change (case v) then there are two possibilities, the SSM is on the maternal copy or the paternal copy.  If the SSM is on the maternal copy, the update rule is:
\begin{align*}
N_i^r &\leftarrow N_i^r + \eta_u C_i^p \ , \\
N_i^v &\leftarrow N_i^v + \eta_u C_i^m \ .
\end{align*}
If however, the SSM is on the maternal copy, the update rule is:
\begin{align*}
N_i^r &\leftarrow N_i^r + \eta_u C_i^m \ , \\
N_i^v &\leftarrow N_i^v + \eta_u C_i^p \ .
\end{align*}
Note that the breakdown of $C_i$ into $C_i^m$ and $C_i^p$ is only required if the CNV occurs after the SSM on the same branch.

Now that we can calculate $N_i^r$ and $N_i^v$, the observation model for the allelic counts has the following generative process (cf. \eq{eqn:pm}):
\begin{equation*}\label{eqn:pm2}
\begin{aligned}
& \mathcal{G}  \sim  \text{TSSB}(\alpha,\gamma,H) \ ; \quad
\tilde{\eta}_{i}  \sim \mathcal{G} \ ; \\
& a_i \mid d_i,\tilde{\eta}_i,\epsilon  \sim \text{Binomial}\left(d_i,\frac{N_i^r (1- \epsilon) + N_i^v \epsilon}{N_i^v + N_i^r}\right) \ .
\end{aligned}
\end{equation*}
Note that in some circumstances, a SSM can be placed on a particular copy of the chromosome by looking for reads that cover the SSM and nearby heterozygous germline mutations. If this is not possible then the likelihood of $a_i$ is the average of two likelihoods; the likelihood if the SSM occurs on the maternal genome and the likelihood if the SSM occurs on the paternal genome. 

\subsection{Extension to multiple samples}
Our model can be easily extended to multiple tumor samples.  We make no assumptions regarding the time that the samples were collected, so this extension is equally applicable to multiple samples collected simultaneously (e.g. as in \cite{Gerlinger12}) or over a period of time as in \cite{Schuh12}.  We allow the tree-structured stick-breaking process prior to be shared across all the samples. The main technical difference between the single and the multiple sample models lies in the sampling procedure for SSM population frequencies. In the multiple sample model, we ensure that the clonal evolutionary constraints are satisfied separately for each tumor sample and then make a global Metropolis-Hastings move based on the product of posterior distributions across all the samples (cf. \eq{eqn:posterior1}).

\subsection{MCMC settings}
In all the MCMC experiments, we fix the number of MCMC iterations to 2,500 and use a burn-in of 100 samples. We also fix the number of iterations in the Metropolis-Hastings algorithm to 5,000 and set the scaling factor for the Dirichlet proposal distribution to $100$ (see PhyloSub paper \cite{phylosub}). We use the CODA R package \cite{coda} for MCMC diagnostics to monitor the convergence of the samplers using the complete-data log likelihood traces and the corresponding autocorrelation function.

\subsection{Sequencing error}
It is becoming increasingly clear that sequencing error is not uniform across the genome and different trinucleotide sequences result in different sequencing error rates \cite{boutros2014global}. While the precise nature of these differences is not yet fully known, PhyloWGS allows the user to input a different sequencing error rate for each mutation.

%%%%%%%%%%%%%%%%%%%%%%%%%%%%%%%%%%%%%%%%%%%%%%%%%%%%%%%%%%%%%%%%

%\section*{Acknowledgements}
  
%%%%%%%%%%%%%%%%%%%%%%%%%%%%%%%%%%%%%%%%%%%%%%%%%%%
\bibliographystyle{unsrt}
\bibliography{biblio}
\end{document}